\documentclass[twocolumn,appendix]{aastex631}

\begin{document}

\title{Central Velocity Dispersion being the Primary Driver of Abundance Patterns in Quenched Galaxies}

\correspondingauthor{Enci Wang, Cheqiu Lyu}
\email{ecwang16@ustc.edu.cn, lyucq@ustc.edu.cn}

\author[0009-0009-2660-1764]{Haixin Li\thanks{Email: 1412893892lhx@gmail.com}}
\affiliation{Department of Astronomy, University of Science and Technology of China, Hefei, Anhui 230026, China}
\affiliation{School of Astronomy and Space Science, University of Science and Technology of China, Hefei, Anhui 230026, China}

\author[0000-0003-1588-9394]{Enci Wang}
\affiliation{Department of Astronomy, University of Science and Technology of China, Hefei, Anhui 230026, China}
\affiliation{School of Astronomy and Space Science, University of Science and Technology of China, Hefei, Anhui 230026, China}

\author[0009-0000-7307-6362]{Cheqiu Lyu}
\affiliation{Department of Astronomy, University of Science and Technology of China, Hefei, Anhui 230026, China}
\affiliation{School of Astronomy and Space Science, University of Science and Technology of China, Hefei, Anhui 230026, China}

\author[0000-0002-4597-5798]{Yangyao Chen}
\affiliation{Department of Astronomy, University of Science and Technology of China, Hefei, Anhui 230026, China}
\affiliation{School of Astronomy and Space Science, University of Science and Technology of China, Hefei, Anhui 230026, China}
\affiliation{Key Laboratory for Research in Galaxies and Cosmology, Department of Astronomy, University of Science and Technology of China, Hefei, Anhui 230026, China}

\author[0000-0002-4911-6990]{Huiyuan Wang}
\affiliation{Department of Astronomy, University of Science and Technology of China, Hefei, Anhui 230026, China}
\affiliation{School of Astronomy and Space Science, University of Science and Technology of China, Hefei, Anhui 230026, China}
\affiliation{Key Laboratory for Research in Galaxies and Cosmology, Department of Astronomy, University of Science and Technology of China, Hefei, Anhui 230026, China}

\author[0009-0004-5989-6005]{Zeyu Chen}
\affiliation{Department of Astronomy, University of Science and Technology of China, Hefei, Anhui 230026, China}
\affiliation{School of Astronomy and Space Science, University of Science and Technology of China, Hefei, Anhui 230026, China}

\author[0009-0008-1319-498X]{Haoran Yu}
\affiliation{Department of Astronomy, University of Science and Technology of China, Hefei, Anhui 230026, China}
\affiliation{School of Astronomy and Space Science, University of Science and Technology of China, Hefei, Anhui 230026, China}

\author[0009-0004-7042-4172]{Cheng Jia}
\affiliation{Department of Astronomy, University of Science and Technology of China, Hefei, Anhui 230026, China}
\affiliation{School of Astronomy and Space Science, University of Science and Technology of China, Hefei, Anhui 230026, China}

\author[0009-0006-7343-8013]{Chengyu Ma}
\affiliation{Department of Astronomy, University of Science and Technology of China, Hefei, Anhui 230026, China}
\affiliation{School of Astronomy and Space Science, University of Science and Technology of China, Hefei, Anhui 230026, China}

%% Note that the \and command from previous versions of AASTeX is now
%% depreciated in this version as it is no longer necessary. AASTeX 
%% automatically takes care of all commas and "and"s between authors names.

%% AASTeX 6.31 has the new \collaboration and \nocollaboration commands to
%% provide the collaboration status of a group of authors. These commands 
%% can be used either before or after the list of corresponding authors. The
%% argument for \collaboration is the collaboration identifier. Authors are
%% encouraged to surround collaboration identifiers with ()s. The 
%% \nocollaboration command takes no argument and exists to indicate that
%% the nearby authors are not part of surrounding collaborations.

%% Mark off the abstract in the ``abstract'' environment. 

\begin{abstract}
The element abundances of galaxies provide crucial insights into their formation and evolution. Using high-resolution IFU data from the MaNGA survey, we analyze the central spectra (0-0.5 $R_{\rm e}$) of 1,185 quenched galaxies ($z = 0.012-0.15$) to study their element abundances and stellar populations. We employ the full-spectrum fitting code {\tt alf} to derive stellar ages and element abundances from synthetic spectra and empirical libraries. Our key findings are: (1) Central velocity dispersion ($\sigma_*$) is the most effective parameter correlating with (relative) element abundances, especially [Na/Fe], [Mg/Fe], [C/Fe], and [N/Fe], outperforming $M_\ast$ and $M_\ast/R_{\rm e}$. (2) When binned by $\sigma_*$, the relative abundances of Na, Mg, C, and N remain stable across different formation times ($T_{\rm form}$), suggesting these elements are primarily influenced by the burstiness of star formation (traced by $\sigma_*$) rather than prolonged evolutionary processes. (3) Fe and Ca show little variation with $\sigma_*$, indicating weaker sensitivity to $\sigma_*$-driven processes. However, $T_{\rm form}$ has a global influence on all elements, contributing to their overall chemical evolution, albeit secondary to $\sigma_*$ for most elements. These results support the primary role of $\sigma_*$ in shaping the abundance patterns, likely stemming from the connection between central massive black holes and possibly dark matter halos, which influences the burstiness of star formation histories.

%while also highlighting the complementary influence of $T_{\rm form}$ on gradual element enrichment.

\end{abstract}

%% Keywords should appear after the \end{abstract} command. 
%% The AAS Journals now uses Unified Astronomy Thesaurus concepts:
%% https://astrothesaurus.org
%% You will be asked to selected these concepts during the submission process
%% but this old "keyword" functionality is maintained in case authors want
%% to include these concepts in their preprints.

%% From the front matter, we move on to the body of the paper.
%% Sections are demarcated by \section and \subsection, respectively.
%% Observe the use of the LaTeX \label
%% command after the \subsection to give a symbolic KEY to the
%% subsection for cross-referencing in a \ref command.
%% You can use LaTeX's \ref and \label commands to keep track of
%% cross-references to sections, equations, tables, and figures.
%% That way, if you change the order of any elements, LaTeX will
%% automatically renumber them.
%%
%% We recommend that authors also use the natbib \citep
%% and \citet commands to identify citations.  The citations are

\section{Introduction} \label{sec:intro}
\setlength{\parindent}{1em}

Metallicity of galaxies provides vital insights into the chemical evolution and formation history of galaxies. Understanding metallicity helps unravel the processes driving galaxy evolution, including star formation, supernova feedback, and gas accretion \citep[e.g.][]{Schaye-10, 2010SpolaorKobayashi+, Bouche-10, Dave-11, Dave-12, Lilly-13, Peng-14, Belfiore-19, Wang-19, Wang-21, Wang-22, 2024ApJ...971L..14M, Wang-24}. While various factors affecting metallicity have been investigated over time, identifying the most influential remains contentious. In the local universe, the well-established mass-metallicity relation (MZR) shows that more massive galaxies tend to be more metal-rich both on stellar metallicity \citep[e.g.,][]{2004ApJ...613..898T,2014ApJ...788...72G,2018ParikhThomas+,2022CarnallMcLure+} and gas metallicity \citep{2024StantonCullen+}. In addition, metallicity also correlates with stellar velocity dispersion ($\sigma_*$) and other structural parameters \citep[e.g.,][]{2005ApJ...621..673T,2014ApJ...780...33C,2015MNRAS.448.3484M, 2024ApJ...971L..14M}.

Aperture effects have posed a persistent challenge in spectroscopic studies, often leading to biased metallicity comparisons across galaxies. These biases stem from the inconsistent aperture sizes used to extract spectra, resulting in inaccuracies, particularly in stacked or comparative analyses of galaxies with varying sizes \citep[e.g.,][]{2005PASP..117..227K}. To address this, \citet{2018MNRAS.479.1807D} proposed aperture-matched subsampling, wherein galaxies are selected with physical sizes matched to the SDSS fiber aperture in units of effective radius ($R_\mathrm{e}$). Their work demonstrated that gas-phase metallicity correlates more tightly with the gravitational potential $(M_*/R_{\rm{e}})$ compared to stellar mass ($M_*$) or surface mass density $(M_*/R^2_{\rm{e}})$, suggesting that potential may be a significant factor in regulating metallicity. While this method reduces aperture bias in traditional single-fiber data, recent advancements in integral field spectroscopy, such as MaNGA \citep[Mapping Nearby Galaxies at Apache Point Observatory;][]{2015ApJ...798....7B} survey, provide a more direct solution. For instance, \citet{2024ApJ...971L..14M} utilized MaNGA data along with TNG50 simulations to explore the relationship between gas-phase metallicity and galaxy properties, avoiding aperture-related biases. They found that galaxy size is more important in determining the gas-phase metallicity of star-forming galaxies than star formation rate (SFR), both in simulations and observations. 
%ir findings showed that the $(M_*/R^{\beta}_{\rm{e}})$ correlates much more strongly with gas-phase metallicity than $M_*$ alone, with the potential index $\beta \sim$ 0.6 $\text{-}$ 1 providing a better fit in both observations and simulations, regardless of star formation rate (SFR).
%With its high spatial resolution and integral field unit (IFU) data \citep{ 2016AJ....152...83L}, MaNGA eliminates the need for aperture-bias corrections and enables precise spectral mapping across different regions of galaxies, such as the central 0$\text{-}$0.5 $R_\mathrm{e}$.

Elliptical galaxies, particularly those that have quenched and ceased star formation, exhibit a complex evolutionary history \citep[e.g.,][]{1987DjorgovskiDavis+,2002ChiosiCarraro+,2005vanDokkum+,2024WangLeja+}. These galaxies typically undergo processes like minor merge and star formation cessation, resulting in diverse and intricate chemical properties \citep{1992NietoBender+,2012AguerriHuertas-Company+,2024Monteiro-OliveiraLin+}. These minor mergers contribute significantly to the increase in both the mass and size of galaxies by accreting smaller satellite galaxies, and this process is largely dependent on environmental conditions \citep{2006DeLuciaSpringel+,10.1093/mnras/stx732,2017ScottBrough+}. This accretion not only adds mass but also causes outward growth, increasing the effective radius \citep{2007A&A...476.1179B,2013MNRAS.429.2924H}. However, this merging behavior after star formation cessation poses a significant challenge in studying the metallicity of quenched galaxies. The accreted material, often from lower-mass and lower-metallicity galaxies, dilutes the metallicity of central regions of galaxies \citep{2007A&A...476.1179B,2019ParikhThomas+}. This dilution complicates interpretations of a galaxy’s formation history, as it is difficult to disentangle the galaxy’s internal evolutionary processes from external merger influences \citep{2020MNRAS.493.6011M,2023ApJ...956L..42S}. As such, the chemical properties of quenched galaxies may reflect the combination of both internal processes and external accretion events.

To reduce the contamination effects of external material and minor mergers, this work focuses on the central regions (0-0.5 $R_\mathrm{e}$) of galaxies, where these influences are less pronounced \citep{2024ZhuangKirby+,2024MoleroMatteucci+}. Although minor merger events after quenching may contaminate some age information, we argue that the central spectra remain relatively reliable for estimating stellar population ages. Focusing on these regions minimizes the influence of recent minor mergers, allowing for a more accurate determination of the formation time. %, which is critical for understanding the metallicity and formation history of galaxies. 
Consequently, age becomes a central parameter in our analysis, alongside key mass-related quantities, such as $\sigma_*$, $M_*$, and $M_*/R_\mathrm{e}$ (gravitational potential). This strategy allows us to explore the relationships between element abundances and different galactic parameters in a more nuanced way.

The relation between $\sigma_*$ and metallicity is particularly significant for quenched galaxies. \citet{2023HongWang+} and \citet{2012WakevanDokkum+} highlighted the significance of $\sigma_*$ in understanding galaxy dynamics, star formation quenching, and black hole growth, which underscores its central role in shaping galaxy evolution. Several studies also proposed that $\sigma_*$ may be a stronger indicator of a galaxy’s chemical enrichment than stellar mass \citep[e.g.,][]{2014ApJ...780...33C,2019ApJ...874...66G}. This view stems from observations that higher $\sigma_*$ correlates with older stellar populations and enhanced $\alpha$-element abundances, suggesting rapid star formation followed by early quenching. However, whether $\sigma_*$ drives these chemical properties alone or in combination with stellar mass remains an ongoing debate.

In quenched galaxies, the $\alpha$-element to iron ratios ([$\alpha$/Fe]), such as [Mg/Fe], often serve as a proxy for star formation timescales. High [$\alpha$/Fe] ratios suggest rapid star formation before Type Ia supernovae could contribute significant amounts of iron. This rapid formation is typically linked with high $\sigma_*$ galaxies, which likely experienced intense early bursts of star formation \citep{2005ApJ...621..673T,2014ApJ...780...33C,2022OyarzunBundy+,2022GuGreene+,2024MatharuNelson+,2024CarnallCullen+}. Such $\alpha$-element enhancements provide insights into the star formation and quenching histories of massive galaxies, revealing their past evolutionary paths. MaNGA’s IFU data facilitates a more refined investigation into the relationship between $\sigma_*$ and metallicity in quenched galaxies.

This work addresses the ongoing debate over whether $\sigma_*$, $M_*$ or $M_*/R_{\rm e}$ is a more reliable indicator of a galaxy’s chemical properties. By utilizing high-precision spectral data and advanced modeling techniques, including element abundance and age data from \citet{2018ApJ...854..139C} model, we aim to clarify the relative importance of these parameters. %By ensuring data integrity and focusing on regions minimally influenced by external factors, we seek to provide new insights into the primary drivers of chemical evolution in galaxies. 
This paper is organized as follows: Section~\ref{sec:2} details the observation data, along with the data processing and modeling methods. Section~\ref{sec:3} presents our results, focusing on the relationships between $\sigma_*$, metallicity, and galaxy age. In Section~\ref{sec:4}, we discuss these findings within the broader context of existing literature and their implications for galaxy formation and evolution. Finally, Section~\ref{sec:5} summarizes our key conclusions and outlines potential avenues for future research.

\section{Data} \label{sec:2}

\subsection{Data Source and Selection}

We employ data from the MaNGA Data Release 17 \citep{2022ApJS..259...35A}, which offers a collection of 2-dimensional spectra from 11,273 galaxies. The data were obtained using the two dual-channel BOSS spectrographs on the Sloan Telescope \citep{2006AJ....131.2332G,2013AJ....146...32S}, covering a wavelength range of 3600–10300 \AA\ with a spectral resolution of approximately 2000. Typically, the spatial coverage of each galaxy extends beyond 1.5 $R_\mathrm{e}$, with a spatial resolution of 1–2 kpc. 

One of the unique strengths of the MaNGA dataset is its inclusion of a considerable number of quenched galaxies because of the flat distribution in stellar mass. While this introduces some selection bias, the large sample helps reduce statistical uncertainty, enabling us to investigate the formation and evolution of low-mass galaxies with improved confidence, ultimately increasing the research value of the dataset. Together with high-mass counterparts, this sample coverage offers a more comprehensive understanding of galaxy properties across a wide mass range.

\begin{figure}[htbp]
    \centering
    \includegraphics[width=0.47\textwidth]{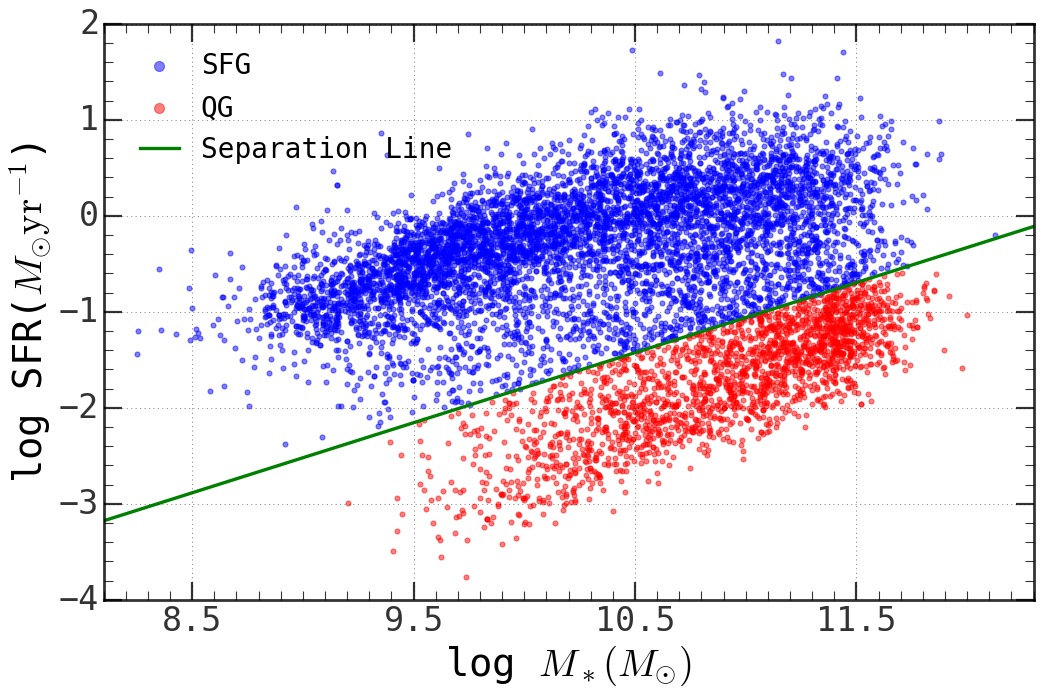}
    \caption{Classification of star-forming galaxies (SFG) and quenched galaxies (QG) based on \hyperref[eq1]{Equation 1}. The blue dots represent star-forming galaxies, while the red dots represent quenched galaxies. The green line indicates the separation line according to \hyperref[eq1]{Equation 1}. }
    \label{Fig1}
\end{figure}

In this work, the stellar mass and total SFR measurements are taken from \cite{2018SalimBoquien+}, derived from spectral energy distribution fittings using GALEX, SDSS, and WISE photometry. To evaluate the importance of the total stellar mass in shaping the central metallicity trends, we opt not to use spaxel-based mass measurements such as those provided by \cite{2022NeumannThomas+} or the Pipe 3D \citep{2022SanchezBarrera-Ballesteros+}. Instead, we use the dataset from \cite{2018SalimBoquien+}, which provides a robust estimate of the total stellar mass for galaxies. To differentiate between star-forming and quenched galaxies, we adopt a boundary based on the SFR$-M_*$ relation as derived in \cite{2016MNRAS.462.2559B}. The method begins with defining the star-forming main sequence (SFMS) by matching galaxies based on their stellar mass and redshift, ensuring that star-forming galaxies are selected by emission line criteria (S/N $\textgreater$ 5, excluding AGN). A logarithmic distance from the star-forming main sequence ($\Delta$SFR) is then calculated, where $\Delta$SFR measures the difference between the galaxy's SFR and the median SFR for galaxies with similar mass and redshift. The threshold for quenched galaxies is set at $\Delta$SFR$\textless-$1, which effectively separates star-forming galaxies from quenched galaxies. To ensure a more conservative selection of quenched galaxies, we modify the original formula by shifting it downward along the y-axis, which appears to be a good demarcation line in Figure \ref{Fig1}, resulting in the following equation:
% In this work, the stellar mass and total SFR measurements are taken from \cite{2014ApJ...797..126S}, derived from spectral energy distribution fittings using GALEX, SDSS, and WISE photometry. We adopt a boundary to differentiate between star-forming and quenched galaxies, which is taken from \cite{2016MNRAS.462.2559B}: 
% \begin{equation}
% \log\left(\text{SFR}/M_{\odot}\,\text{yr}^{-1}\right) = 0.73 \times \log(M_*/M_{\odot}) - 8.09.
% \label{eq1}
% \end{equation}
% To adopt a more conservative and pure selection for quenched galaxies, we modify the equation from \cite{2016MNRAS.462.2559B} by subtracting 1 dex, resulting in following equation:
\begin{equation}
\log\left(\text{SFR}/M_{\odot}\,\text{yr}^{-1}\right) = 0.73 \times \log(M_*/M_{\odot}) - 9.09.
\label{eq1}
\end{equation}
Following this criterion, we selected 1,499 quenched galaxies with redshifts from 0.012 to 0.15. The integral field unit (IFU) data from MaNGA alleviates these concerns by providing spatially resolved spectra, allowing us to measure metallicity across the entire central region rather than from a single aperture. To ensure consistency in our analysis, we focused on the spectra from the central regions within $R/R_{\rm e}$ = 0–0.5 for all galaxies. This proportional selection of central regions, regardless of galaxy size, enables uniform comparisons across the entire sample, allowing us to accurately study the central properties of galaxies consistently.

\subsection{Data Processing}

The central spectra of these 1,499 galaxies are subjected to stellar velocity correction (based on the output from the MaNGA data-analysis pipeline) and interpolation, allowing the spectra to be analyzed in their rest-frame wavelengths. To improve spectral quality and reduce the influence of noise and outliers, we employ a stacking method by summing the spexal spectra within 0.5$R_{\rm e}$ for each galaxy. %, which enhances the signal-to-noise ratio (SNR) of spectra.  %, providing a clearer and more reliable representation of each galaxy’s central spectral properties. 
The stacked spectra allow us to accurately capture the central spectral characteristics of the galaxies, effectively minimizing the impact of outliers and random noise, which ensures robust analysis of their central regions.

If the number of spaxels within 0.5 $R_\mathrm{e}$ regions for one galaxy is $N_{\rm b}$,  %, which is crucial for ensuring sufficient spatial resolution in our analysis. Using this pixel information
we calculate the signal-to-noise ratio (SNR) in the following way, accounting for both noncovariant and covariant errors based on MaNGA guidelines \citep{2016AJ....152...83L,2019AJ....158..231W}. The ratio of these errors is given by: 
\begin{equation}
n_{\rm covar}/n_{\rm nocovar} = 1 + 1.62 \log(N_{\rm b}) \quad \text{for} \quad N_{\rm b} < 100,
\label{eq2}
\end{equation}
and 
\begin{equation}
n_{\rm covar}/n_{\rm nocovar} = 4.2 \quad \text{for} \quad N_{\rm b} \geq 100,
\label{eq3}
\end{equation}
Here, $n_{\rm nocovar}$ represents the error after binning without considering the covariance between the binned spaxels based on the inverse variance of the datacubes. %This type of error assumes that each spaxel is independent of the others, with no correlation between their respective uncertainties.
While $n_{\rm covar}$ is the error that accounts for covariant, where correlations exist between spaxels, which can occur due to processes like interpolation or smoothing during data reduction.

To quantify the relationship between these errors, the ratio $n_{\rm covar}/n_{\rm nocovar}$ is given by Equation \ref{eq2} and Equation \ref{eq3}. For a small number of binned spaxels ($N_b < 100$), the covariant error scales with the number of spaxels as $1 + 1.62 \log(N_b)$, reflecting the increased error due to correlations as more spaxels are binned together. For larger numbers of spaxels ($N_b \geq 100$), this ratio becomes a constant 4.2, indicating that beyond a certain threshold, additional binning has a less significant impact on increasing the covariance of the data.

\begin{figure}
    \centering
    \includegraphics[width=0.47\textwidth]{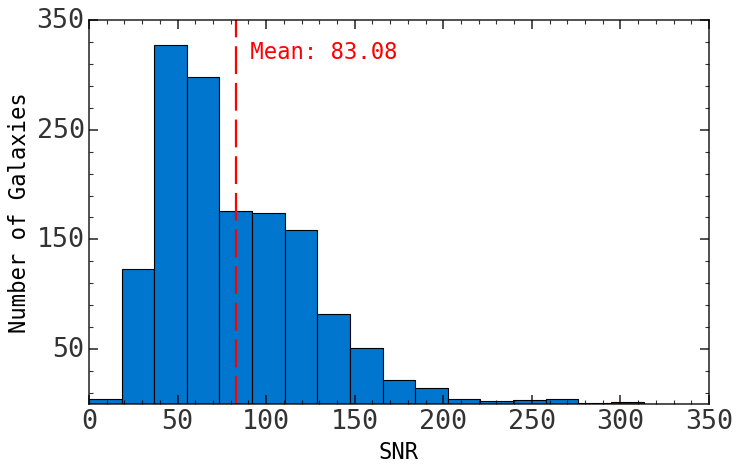}
    \caption{Signal-to-noise ratio (SNR) distribution of the stacked spectra for all galaxies in our sample. Based on the calculation method from Equations \ref{eq2} and \ref{eq3}, there are 1,499 galaxies with valid SNR values. The stacked spectra show high quality, with an average SNR of 83.08.}
    \label{Fig2}
\end{figure}

This stacking method significantly enhances the SNR of spectra. Figure~\ref{Fig2} shows the SNR distribution across the sample, which yields a mean SNR value of 83.08. Such a high SNR demonstrates the clarity and reliability of the central spectral properties, enabling the possibility of obtaining element abundances for individual galaxies. 
%rather than relying on bin-stacking methods.

Some spectra, particularly those at the higher wavelength end (above 8,700 \r{A}), exhibit poor quality due to skyline contamination, leading to significant data fluctuations. To address this, %we conduct outlier detection on the tail-end of the spectra, excising portions of data where extreme values were identified. 
we narrow our analysis to the wavelength range of 3,900--9,100 \r{A}.  This not only reduces the impact of tail-end contamination but also encompasses the key absorption lines required by our modeling program.

After processing, we generate stacked spectra for each galaxy by combining individual spexal spectra within the central regions. One example of the stacked spectra is shown in Figure~\ref{Fig4}, demonstrating the robustness and clarity of our spectral extraction method.

\begin{figure}
    \centering
    \includegraphics[width=0.47\textwidth]{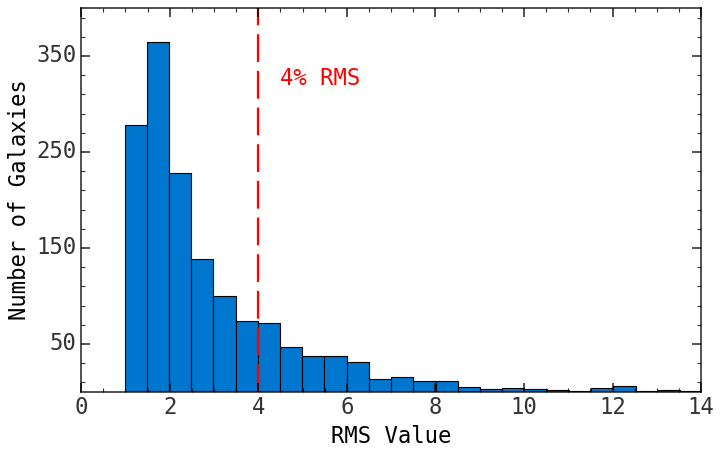}
    \caption{Root-mean-square (RMS) distribution of the fitted spectra. The RMS values are mainly concentrated in the range of 1-2.5. For data validity, spectra with RMS values above 4 are cut, resulting in a final sample of 1,185 galaxies.}
    \label{Fig3}
\end{figure}

\subsection{Spectral Fitting}
We utilized a custom full-spectrum fitting code, {\tt alf} \citep{2018ApJ...854..139C}, based on the model from \cite{2012ConroyvanDokkum+}, to measure individual element abundances and stellar population ages. The {\tt alf} models, designed for older stellar populations ($>$1 Gyr) across a wide range of metallicities ($-1.5 \lesssim$ [Fe/H] $\lesssim 0.3$), predict spectral variations as a function of both metallicity and age. Covering the optical–NIR range (0.37–2.4 $\mu$m), these models allow for variations in 18 individual elements, including C, N, O, Mg, Si, Ca, Ti, and Fe, and have been validated against integrated light spectra from Galactic globular clusters with well-established stellar population parameters. The models integrate metallicity-dependent MIST isochrones \citep{2016ApJ...823..102C}, empirical spectral libraries from MILES and IRTF \citep{2006A&A...457..787S,2017ApJS..230...23V}, synthetic spectra, offering theoretical response functions based on two initial mass function (IMF) modes: Kroupa \citep{2001MNRAS.322..231K} and Salpeter \citep{1955Salpeter+}. For our analysis, we adopt the response function based on Kroupa IMF. The fitting process involves optimizing 46 free parameters, including velocity offset, low-mass IMF slope, velocity dispersion, SSP-equivalent stellar population age, isochrone metallicity, and the abundances of key elements. Additional parameters account for Balmer emission line flux, velocity, and broadening of emission lines, shifts in the effective temperature of the fiducial isochrones ($T_{\rm eff}$), and an instrumental jitter term to adjust for uncertainties. By utilizing full-spectrum fitting, the models significantly reduce uncertainties compared to traditional line index fitting methods \citep[e.g.][]{2012JohanssonThomas+, Thomas2010}. During the fitting, the IMF results show a slope range from 2.2 to 2.7, for lower $T_{\text{form}}$ to higher $T_{\text{form}}$. The values are consistent with the results in \cite{2018ParikhThomas+}.

We fit each of the 1,499 stacked spectra individually using {\tt alf}. To ensure the accuracy of the fitted data, we filter out results with an RMS error greater than 4\%. The RMS distribution of the fitted spectra is shown in Figure.~\ref{Fig3}, where the final set of selected galaxies demonstrates consistent fitting results. This filtering process yielded a final sample of 1,185 galaxies, providing a robust dataset for investigating the chemical properties and stellar populations of quenched galaxies.

\begin{figure*}[htbp]
    \centering
    \includegraphics[width=1\textwidth]{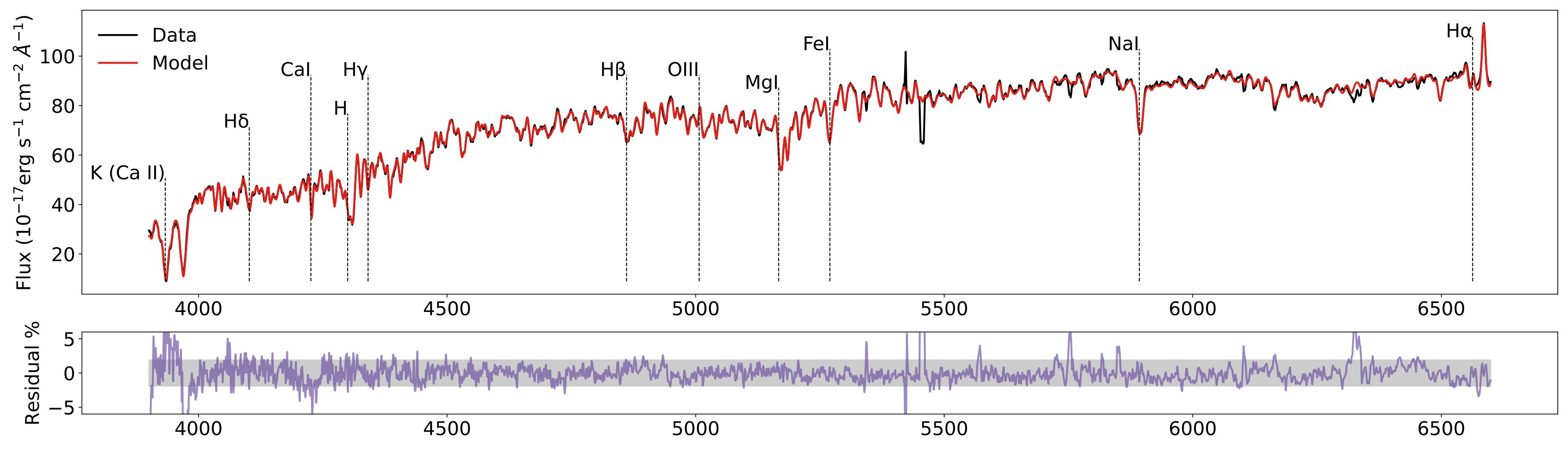}   
    \includegraphics[width=1\textwidth]{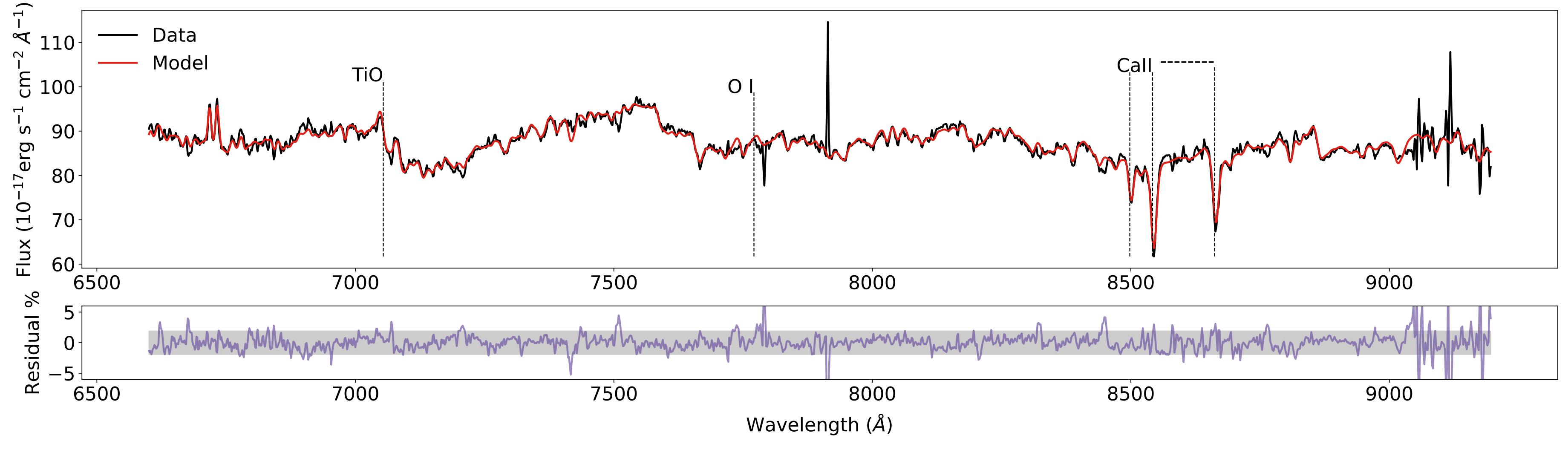}   
    \caption{Example of original and fitted spectra for a galaxy (plate-ifu ID: 8077-3701) with a median RMS value (1.9122) in our sample. The black, red, and purple lines represent the original spectrum, fitted spectrum, and residual, respectively. The shaded regions indicate the RMS level and the spectra are divided into two segments for easier display.}
    \label{Fig4}
\end{figure*}

\section{Results and Comparison} \label{sec:3}

\begin{figure*}[htbp]
    \centering
    \includegraphics[width=1\textwidth]{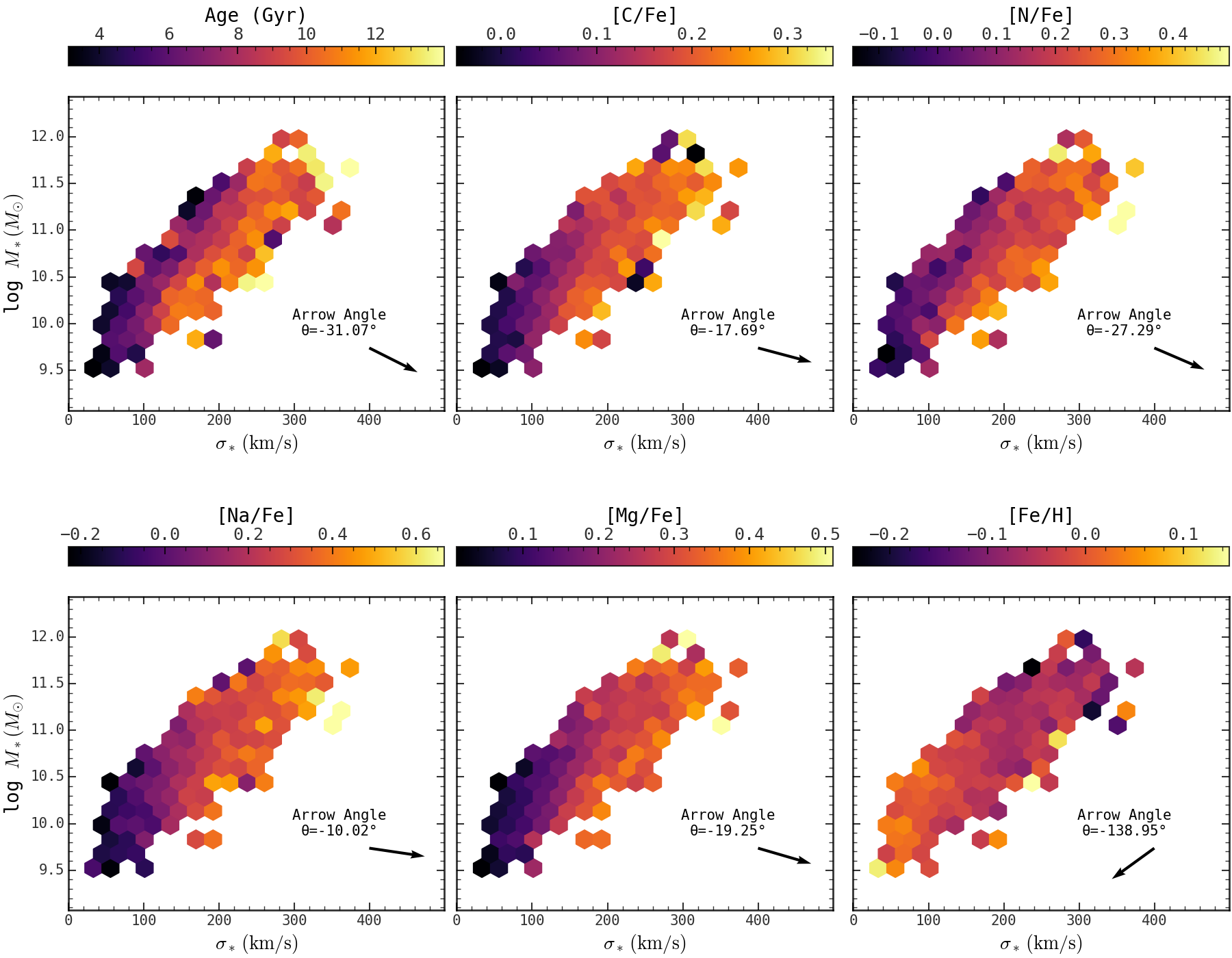}
    \caption{
    Stellar age and (relative) element abundances [C/Fe], [N/Fe], [Na/Fe], [Mg/Fe], and [Fe/H] as functions of velocity dispersion \(\sigma_*\) and stellar mass \(M_*\). The color gradients, indicated by arrows, represent the directions of increasing abundance or age. The trends for stellar age, [C/Fe], [N/Fe], [Na/Fe], and [Mg/Fe] show strong positive correlations with \(\sigma_*\). In contrast, [Fe/H] exhibits a distinct trend, aligning with \(M_*\) but showing an inverse correlation with \(\sigma_*\), opposite to the other elements. [Ca/Fe] is not shown here due to its high stability in the \(M_*\)-\(\sigma_*\) plane.
    }
    \label{Fig5}
\end{figure*}

\begin{figure*}[htbp]
    \centering
    \includegraphics[width=1\textwidth]{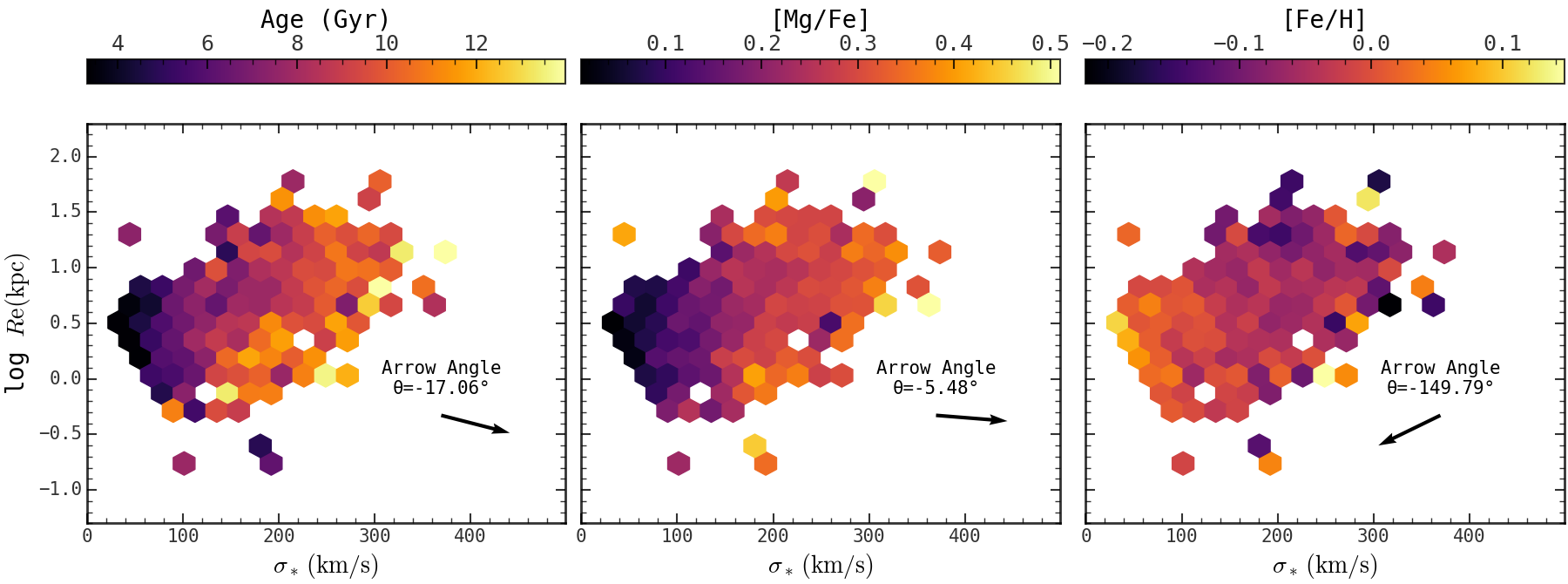}
    \vspace*{0.015\textwidth}
    \\
    \includegraphics[width=1\textwidth]{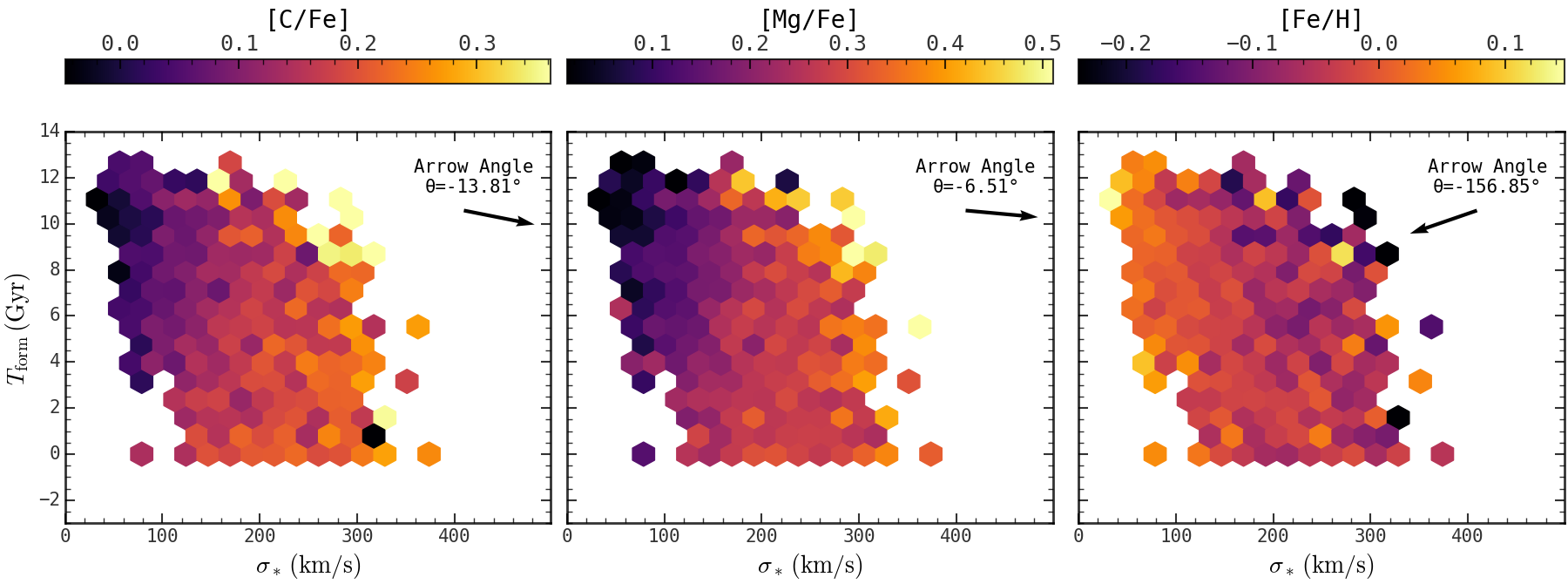}
    \caption{
    (Relative) element abundances [C/Fe], [Mg/Fe], and [Fe/H] as functions of \(\sigma_*\) with \(R_\mathrm{e}\) (top panels) and \(T_{\text{form}}\) (bottom panels) as the y-axis. In the top left panel, [C/Fe] is replaced by age. In each panel, the near-horizontal arrow suggests that both \(R_\mathrm{e}\) and \(T_{\text{form}}\) exhibit weaker correlations with age or (relative) element abundances compared to \(\sigma_*\).
    }
    \label{Fig6}
\end{figure*}

In this section, we present the findings obtained from our {\tt alf} fitting and analysis, with a specific focus on the (relative) element abundances of Na, Mg, Ca, Fe, C, and N. These elements are selected as they represent key processes in the chemical evolution of galaxies and provide a comprehensive view of their star formation history and nucleosynthesis.

Magnesium (Mg), Sodium (Na), and Calcium (Ca) are all closely related in terms of their nucleosynthetic origins, despite some differences. Mg is a classic $\alpha$-element, predominantly produced in Type II supernovae (SNe) during the core-collapse of massive stars during the early stages of a galaxy's evolution, serving as a reliable tracer of early star formation due to its production in short-lived, massive stars \citep[e.g.,][]{2007A&A...465..799L}. Na, although not a strict $\alpha$-element, behaves similarly to Mg in many galactic environments due to its production in core-collapse SNe and its abundance being linked to Mg through nucleosynthetic pathways like the Ne-Na cycle \citep[e.g.,][]{2010A&A...509A..88A}. Calcium (Ca), another $\alpha$-element, has a more complex formation history; while it is partially produced in Type II SNe like Mg and Na, a significant portion of Ca is synthesized in Type Ia SNe, contributing later in a galaxy's life \citep[e.g.,][]{2013ARA&A..51..457N}. This dual origin makes Ca an essential element for tracing both early and late chemical enrichment processes.

Iron (Fe), in contrast, is mainly produced in Type Ia SNe, which occur on longer timescales than in Type II events. The delayed enrichment from these SNe increases Fe abundance over time, making Fe a standard measure of galactic metallicity and long-term star formation \citep[e.g.,][]{1997ARA&A..35..503M}.

Carbon (C) and Nitrogen (N) come from different stellar processes. C is produced in the later stages of stellar evolution, especially in low- to intermediate-mass stars during the AGB phase, while N is primarily synthesized through the CNO cycle in intermediate-mass stars \citep[e.g.,][]{2002ApJ...577..281C}. Their abundances reflect contributions from longer-lived stars, complementing the information provided by $\alpha$-elements and Fe.

By analyzing the abundances of these six elements, we obtain a multi-dimensional perspective on the chemical enrichment of galaxies, spanning from early-rapid star formation events to the long-term contributions from both high- and low-mass stars. This comprehensive approach enables us to trace the complex history of star formation, nucleosynthesis, and chemical evolution in galaxies.

As discussed in the introduction, even nearby quenched galaxies comprise both older and younger stellar populations, and these populations can be further complicated by the effects of past merger events. However, by concentrating on the central spectra (0-0.5 $R{\rm e}$) of these galaxies, we can effectively minimize the impact of external material introduced through mergers, providing a more accurate reflection of the stellar population at the time of formation. %This targeted approach allows us to derive reliable age data from our spectral fitting, which we use as a key parameter in our analysis.

Previous studies %, such as those by \citet{2018MNRAS.479.1807D,2024ApJ...971L..14M} and \citet{2018BaroneD'Eugenio+}, 
have highlighted the importance of potential (indicated as $M_*/R_\mathrm{e}$) in determining the metallicity of galaxies \citep[e.g.,][]{2018BaroneD'Eugenio+, 2018MNRAS.479.1807D,2024ApJ...971L..14M}. Their results suggest that potential plays a critical role in regulating the chemical composition of galaxies. Motivated by these findings, we decide to incorporate potential alongside stellar mass and central velocity dispersion as key parameters to explore their comparative influence on the metallicity of galaxy centers. By leveraging this age data alongside these mass-related quantities, we can investigate the relationships between these physical properties and the element abundances. %This method enables us to discern how different elements correlate with galaxy formation and evolution processes, enhancing our understanding of the complex interplay between these parameters in shaping the chemical characteristics of galaxies.

\begin{figure*}[htbp]
    \centering
    \includegraphics[width=1\textwidth]{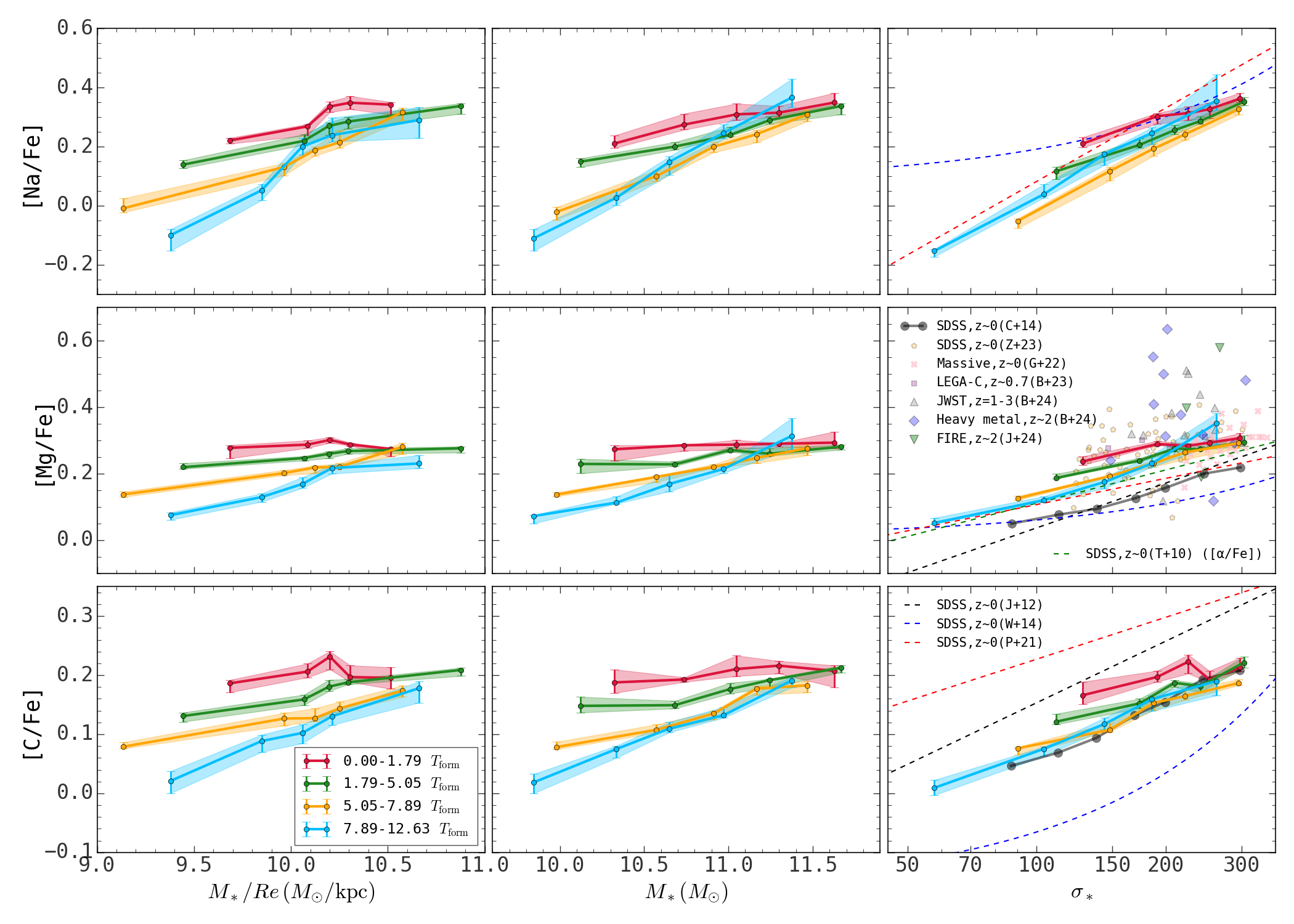}
    \caption{Relative abundances [Na/Fe], [Mg/Fe], and [C/Fe] as a function of $M_*/R_{\rm e}$ (left panels), $M_*$ (middle panels), and $\sigma_*$ (right panels). Lines of various colors represent different $T_{\rm form}$ (in Gyr), with shaded areas denoting 1$\sigma$ uncertainties. Markers represent datasets from SDSS, $z \sim 0$ \citep[e.g.,][(T+10), (J+12), (W+14), (C+14), (P+21) and (Z+23)]{Thomas2010,2012JohanssonThomas+, 2014WortheyTang+, 2014ApJ...780...33C, 2021ParikhThomas+, 2023ApJ...948..132Z}, Massive, $z \sim 0$ \citep[][]{2022ApJ...932..103G}, LEGA-C, $z \sim 0.7$ \citep[][]{2023ApJ...948..140B}, JWST, $z \sim 1-3$ \citep[][]{2024arXiv240702556B}, Heavy Metal, $z \sim 2$ \citep[][]{2024ApJ...966..234B}, and FIRE, $z=2$ \citep[][]{2024arXiv240603549J}. Our results for [Na/Fe] and [C/Fe] agree with previous SDSS-based studies \citep[e.g.,][]{2012JohanssonThomas+,2014WortheyTang+,2021ParikhThomas+}, and our [Mg/Fe] values also align with earlier low-redshift studies at $z \sim 0$ and $z \sim 0.7$ \citep[e.g.,][]{Thomas2010,conroy2013early,2022ApJ...932..103G,2023ApJ...948..132Z,2023ApJ...948..140B}. Higher-redshift data \citep[][FIRE, $z \sim 2$]{2024arXiv240702556B,2024ApJ...966..234B,2024arXiv240603549J} shows increased [Mg/Fe] at large $\sigma_*$, suggesting enhanced $\alpha$-element production likely due to differences in star formation timescales across cosmic time.}
    \label{Fig7}
\end{figure*}

As shown in Figure ~\ref{Fig5}, we present the results of our {\tt alf} fitting for the stellar age and the (relative) abundances of five elements ([Na/Fe], [Mg/Fe], [Fe/H], [C/Fe], [N/Fe]) as a function of $\sigma_*$ and $M_*$. The color coding in the plot reflects the values of each property. In this work, we introduce an angle metric to quantify the relative contributions of $\sigma_*$ and $M_*$ to the element abundances and stellar ages. This angle, calculated using the partial correlation method from \citet{2024ApJ...961..163B}, represents the optimal path through a two-dimensional parameter space. An angle of 0° indicates a full dependence on $\sigma_*$, while 90° indicates full dependence on $M_*$. An angle of 45° suggests an equal contribution from both parameters. 

It is evident from the arrow angle that Age, [Na/Fe], [Mg/Fe], [C/Fe], and [N/Fe] follow a similar trend. Specifically, for a given $M_*$, higher $\sigma_*$ corresponds to higher relative
abundances and older ages, while for a given $\sigma_*$, an increase in $M_*$ leads to lower relative abundances and younger ages. The arrow angle clearly highlights the dominant influence of $\sigma_*$ over $M_*$ in driving these trends. In contrast, although [Fe/H] shows a similar trend with $M_*$ as the other elements, its dependence on $\sigma_*$ is entirely opposite, with lower $\sigma_*$ values corresponding to higher [Fe/H] abundances. On the other hand, [Ca/Fe] shows no significant trend and is therefore not highlighted in the figure. The full set of element trends can be viewed in the Appendix.

In Figure~\ref{Fig6}, we replace the y-axis parameter $M_*$ with effective radius ($R_\mathrm{e}$) and formation time ($T_{\rm form}$) to further explore their relationships with element abundances. Here, $T_{\rm form}$ refers to the formation time of galaxies, calculated by subtracting the stellar age (log age) values derived from our {\tt alf} fitting from the total age of the universe, assumed to be 13.8 billion years \citep{2016A&A...594A..13P}. The arrow angle reveals a stronger dominance, with the angles clustering closer to 0° or 180°, indicating a clearer distinction between the dependencies on $\sigma_*$ and these parameters. This suggests that $\sigma_*$ plays a more significant role in shaping the chemical properties when compared to $R_\mathrm{e}$ and $T_{\rm form}$, further supporting the conclusions drawn from the angle analysis.

\subsection{Dependence on \(\sigma_*\), \(M_*\), and \(M_*/R_\mathrm{e}\)}

\begin{figure*}[htbp]
    \centering
    \includegraphics[width=1\textwidth]{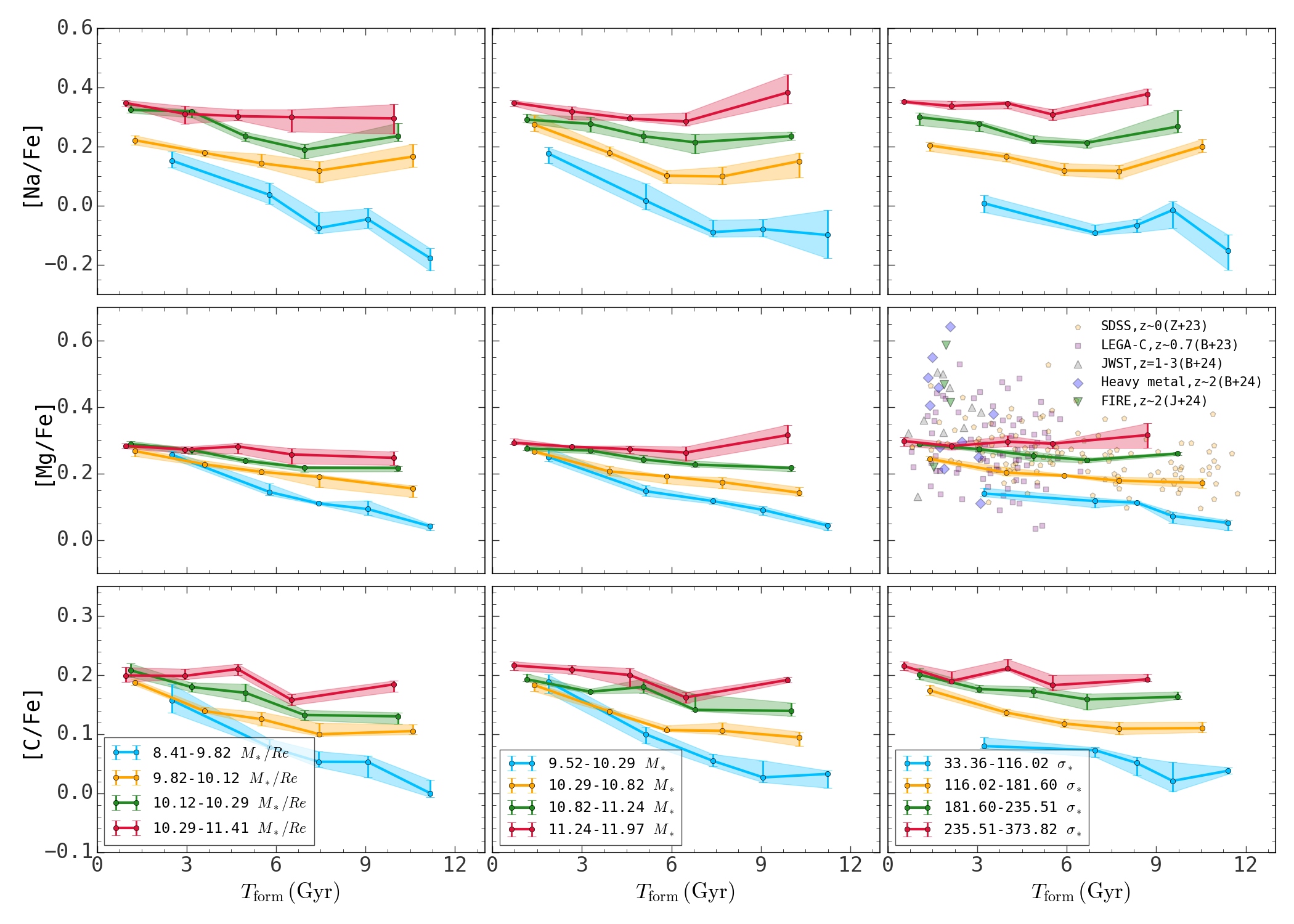}
    \caption{Relative abundances [Na/Fe], [Mg/Fe], and [C/Fe] as a function of $T_{\rm form}$. Lines of various colors represent different $M_*/R_{\rm e}$ (left panels, in $\log(M_{\odot}/\mathrm{kpc})$), $M_*$ (middle panels, in $\log(M_{\odot})$), and $\sigma_*$ (right panels, in km/s), with shaded areas denoting 1$\sigma$ uncertainties. Data points follow the same legend as \hyperref[Fig7]{Figure 7}. The [Mg/Fe] and [Fe/H] abundances are consistent with other low-redshift studies \citep[e.g.,][]{2023ApJ...948..132Z,2023ApJ...948..140B}. Higher-redshift results \citep{2024arXiv240702556B,2024ApJ...966..234B,2024arXiv240603549J} show increasing [Mg/Fe] at lower $T_{\rm form}$, indicating shifts in chemical evolution processes and possibly reflecting changes in star formation history over cosmic time.}
    \label{Fig8}
\end{figure*}

We begin by comparing the effectiveness of different mass-related quantities—velocity dispersion, stellar mass, and mass-to-effective radius ratio — as indicators of element abundance trends across the galaxy sample with $T_{\rm form}$. Figure~\ref{Fig7} presents the relationship between these parameters and the element abundances across different $T_{\rm form}$ bins. To systematically explore these relationships, we employ a percentile-based binning method for all three physical quantities. Given that the MaNGA data is primarily concentrated in the intermediate range of values, we used the 0, 15, 45, 75, and 100 percentiles as binning boundaries, with the 15th, 45th, and 75th percentiles dividing the data into four equal-sized bins. These bins are represented by the colors red, yellow, blue, and green, respectively. Within each bin, we further divide the data into five sub-bins with nearly equal number of data points, and compute the median values for each sub-bin. The uncertainties were obtained through a bootstrapping process after binning and are represented as shaded regions of the corresponding colors in the figure.

As shown in the $\sigma_*$ column, the trends exhibit a notably stronger positive correlation with the relative abundances of [Mg/Fe], [Na/Fe], [C/Fe], and [N/Fe] compared to $M_*$ and $M_*/R_\mathrm{e}$, both of which display flatter trends and a broader range of abundance levels. In addition to the stronger correlation, the $\sigma_*$ column shows the smallest dispersion across the different $T_{\rm form}$ bins, highlighting its closer relationship to the central galactic metallicity. Moreover, aside from the youngest $T_{\rm form}$ bin (0-1.79 Gyr), the remaining three $T_{\rm form}$ bins in the $\sigma_*$ column exhibit high consistency, with the data points aligning closely along a well-defined sequence. These results indicate that $\sigma_*$ is a better indicator of galaxies element abundances compared to $M_*$ and $M_*/R_\mathrm{e}$. Additionally, the consistency observed across different formation times, particularly the minimal impact of $T_{\rm form}$ on abundances in the $\sigma_*$ column, suggests that formation time plays a limited role in shaping the abundance patterns once $\sigma_*$ is considered.

To better assess the impact of $T_{\rm form}$ on overall abundance trends, we adjust our binning and coordinate system. Specifically, we swap the roles of $M_*/R_\mathrm{e}$, $M_*$, and $\sigma_*$ as the binning parameters, using $T_{\rm form}$ as the x-axis. This allows for a more direct evaluation of how element abundances evolve across different formation times under the influence of these mass-related quantities. As shown in Figure~\ref{Fig8}, when binned by $\sigma_*$, the element abundances exhibit minimal correlation with $T_{\rm form}$, displaying nearly parallel abundance lines with weak or no slope, while the $M_*/R_\mathrm{e}$ and $M_*$ columns both show a more obvious declining trend as the $T_{\rm form}$ increase. This suggests that the element abundance levels are strongly tied to $\sigma_*$, with minimal interference from age-related variations. In contrast, $M_*$ and $M_*/R_\mathrm{e}$ show stronger correlations with $T_{\rm form}$, particularly at lower formation times, indicating a greater coupling between these mass-related quantities and the age of the stellar populations.

\begin{figure*}
    \centering
    \includegraphics[width=1\textwidth]{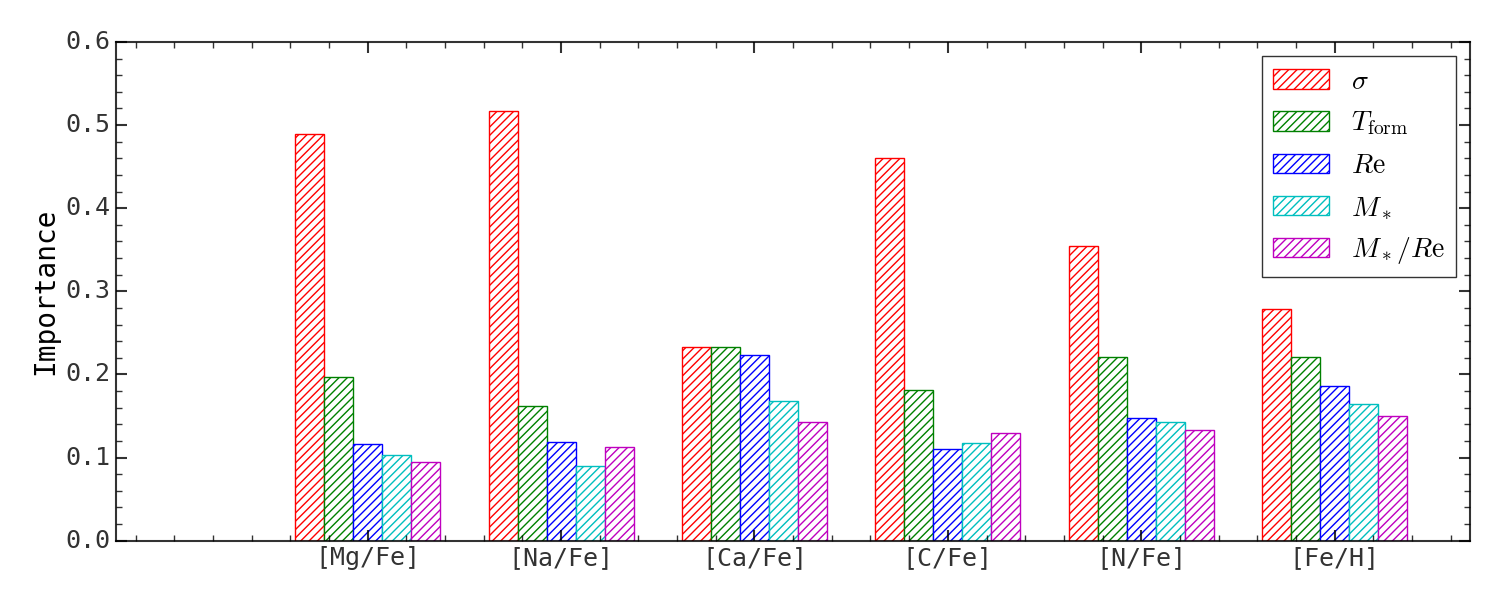}
    \caption{Feature importance of all input properties (labeled in the legend) employed for regressing various relative element abundances
    (labeled on the x-axis) in Random Forest models.}
    \label{Fig9}
\end{figure*}
To quantify these relationships more rigorously, we employ a Random Forest regression analysis using the \texttt{RandomForestRegressor} class from the \texttt{sklearn.ensemble} library. We specify the model with \texttt{n\_estimators=100}, setting the number of trees in the forest to 100 to balance between performance and computational efficiency. Other hyperparameters, such as \texttt{max\_depth} and \texttt{min\_samples\_split}, are left at their default values to allow the algorithm to fully grow each tree and adaptively determine splits. As illustrated in Figure~\ref{Fig9}, the Random Forest regression analysis reveals that $\sigma_*$ plays a dominant role in predicting the relative element abundances, particularly for [Mg/Fe], [Na/Fe], and [C/Fe]. The importance of $\sigma_*$ in these elements far exceeds that of the other parameters, effectively governing the abundance patterns. For [N/Fe], although $\sigma_*$ is still the most influential factor, the difference is less pronounced. In contrast, for [Fe/H], $\sigma_*$ only slightly edges out the other parameters, and for [Ca/Fe], $\sigma_*$ shows almost no leading influence, with $T_{\rm form}$ and $M_*/R_\mathrm{e}$ ranking similarly in predictive power.

Notably, $T_{\rm form}$ consistently ranks as the second most important parameter for all elements after $\sigma_*$, indicating that while $\sigma_*$ is the primary driver, $T_{\rm form}$ still plays a meaningful role in shaping element abundances. This suggests that $T_{\rm form}$ contributes to the overall enrichment history, but its influence is secondary to the strong correlation observed with $\sigma_*$.

Theoretical interpretation of these findings, when combined with the trends observed in Figures~\ref{Fig7} and \ref{Fig8}, reinforces the idea that $\sigma_*$ plays a critical role in determining the relative abundances of [Mg/Fe], [Na/Fe], [C/Fe], and [N/Fe]. The dominance of $\sigma_*$ for these elements suggests that their enrichment is primarily driven by the sharpness and burstiness of the star formation history (SFH), characterized by rapid and intense star formation period \citep[e.g.,][]{2005ApJ...621..673T, Thomas2010}. Although $T_{\rm form}$ contributes to the overall abundance levels, its role is secondary to the effects of star formation burstiness that are closely tied to $\sigma_*$. In contrast, for elements like Fe and Ca, where the correlation with $\sigma_*$ is weaker or comparable to $T_{\rm form}$, the results suggest that their enrichment is more dependent on the long-term, secular evolution tied to $T_{\rm form}$. This implies that the (relative) abundances [Fe/H] and [Ca/Fe] may be governed more by steady, prolonged star formation, rather than the short star formation bursts that primarily influence the other elements. As such, $T_{\rm form}$ plays a more prominent role in the chemical evolution of these elements, reflecting a different enrichment history compared to [Mg/Fe], [Na/Fe], [C/Fe], and [N/Fe].

In light of these results, we conclude that $\sigma_*$ is the most effective mass-related parameter for analyzing metallicity trends in galaxies. Moving forward, we plan to extend this analysis to different radial regions of galaxies to further explore the impact of minor mergers, particularly in the outer regions, to validate our hypothesis that minor mergers have a limited influence on $\sigma_*$ compared to their effects on $M_*$ and the associated element abundances.

\subsection{Analysis on $\sigma_*$ and $T_{\rm form}$}

Having established $\sigma_*$ as the most critical mass-related parameter influencing the element abundances in galaxies, we now proceed to a detailed analysis that incorporates both $\sigma_*$ and stellar age, represented by the $T_{\rm form}$. In the following sections, we focus on examining the relationship between $\sigma_*$ and $T_{\rm form}$ across all six elements (Na, Mg, Ca, Fe, C, and N) in detail to better understand how these two fundamental quantities shape the chemical evolution of galaxies. This approach allows us to compare our results with data from previous studies at different redshifts and discuss them in the context of other surveys and projects.

\begin{figure*}
    \centering
    \includegraphics[width=0.32\textwidth]{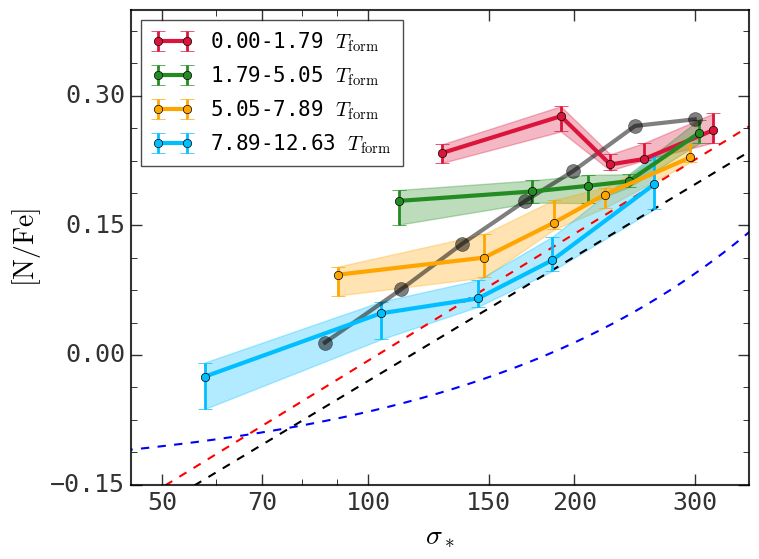}
    \includegraphics[width=0.32\textwidth]{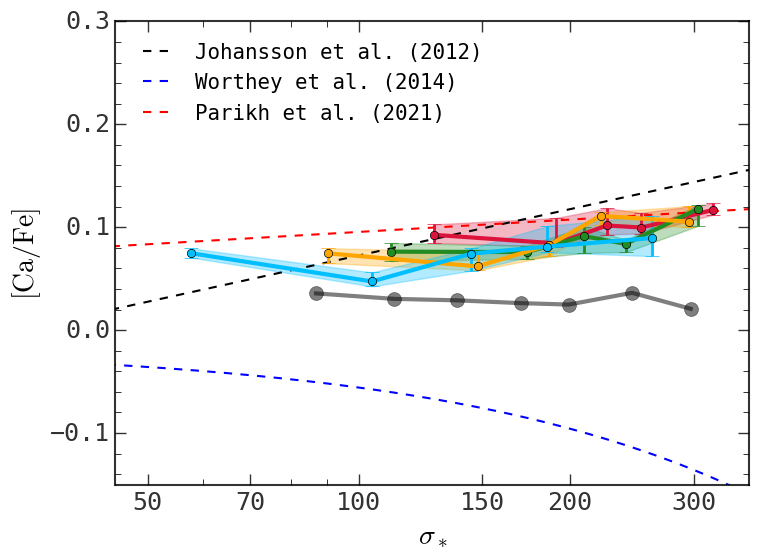}
    \includegraphics[width=0.32\textwidth]{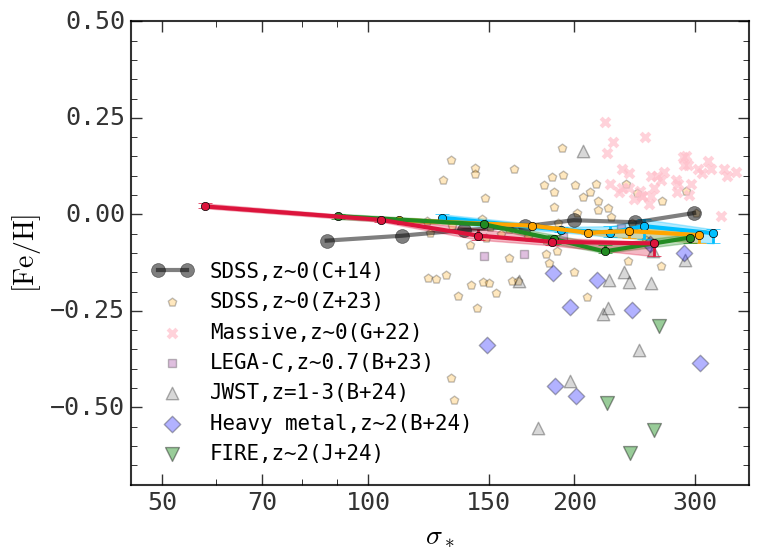}
    \caption{(Relative) Element abundances as a function of $\sigma_*$, binned by $T_{\rm form}$ (in Gyr). Data points follow the same legend as Figure \ref{Fig7}. Our results for [N/Fe] are slightly higher than those reported in previous low-redshift studies \citep[e.g.,][]{2012JohanssonThomas+,2014WortheyTang+,2021ParikhThomas+}, while our result for [Ca/Fe] is consistent with \cite{2012JohanssonThomas+,2021ParikhThomas+}, but do not follow the decreasing trend observed in \cite{2014WortheyTang+}.} Similar to [Mg/Fe], our results in [Fe/H]-$\sigma_*$ plane are consistent with previous studies at $z \sim 0$ and $z \sim 0.7$ \citep[e.g.,][]{2023ApJ...948..132Z,2023ApJ...948..140B}. At higher redshifts \citep{2024arXiv240702556B,2024ApJ...966..234B,2024arXiv240603549J}, [Fe/H] decreases at high $\sigma_*$, suggesting iron enrichment suppression in high-dispersion environments.
    \label{Fig10}
\end{figure*}

Figure~\ref{Fig10} and the third column of Figure~\ref{Fig7} reveal that, when binned by $T_{\rm form}$, the relative abundances [N/Fe], [Na/Fe], [Mg/Fe], and [C/Fe] exhibit similar increasing trends with $\sigma_*$. The different $T_{\rm form}$ bins display nearly identical slopes and abundance levels, indicating a consistent relationship across formation times for these four elements. Notably, all elements not only exhibit small dispersions but also converge to similar abundance levels at higher $\sigma_*$ values. Specifically, [Na/Fe] increases from $-$0.22 to 0.4, [Mg/Fe] ranges from 0.02 to 0.34, [C/Fe] increases from 0.0 to 0.2, and [N/Fe] ranges from $-$0.07 to 0.26. Among the four elements, Na experiences the most dramatic enhancement, while C shows the smallest increase.

In contrast, Figure~\ref{Fig10} also demonstrates that [Ca/Fe] and [Fe/H] show minimal variation in (relative) abundances with $T_{\rm form}$, regardless of $\sigma_*$ binning. For [Ca/Fe], the relative abundances are nearly identical across different $\sigma_*$ bins, indicating a lack of dependence on $\sigma_*$. However, [Fe/H] displays a slight inverse trend, where lower $\sigma_*$ values are associated with higher [Fe/H] abundances, which contrasts with the positive trends observed for the other elements.

\begin{figure*}[htbp]
    \centering
    \includegraphics[width=0.32\textwidth]{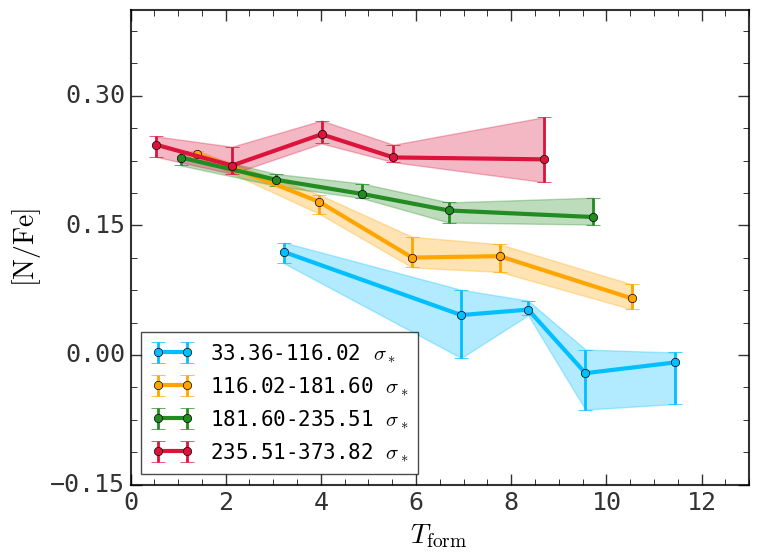}
    \includegraphics[width=0.32\textwidth]{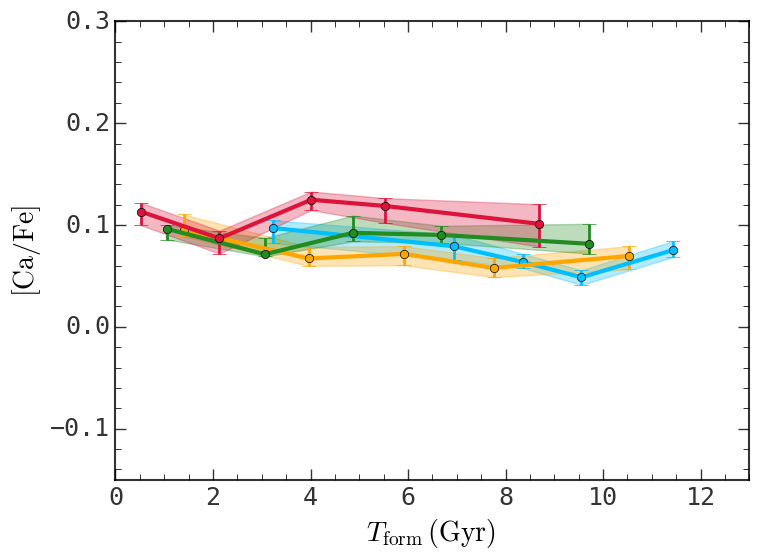}
    \includegraphics[width=0.32\textwidth]{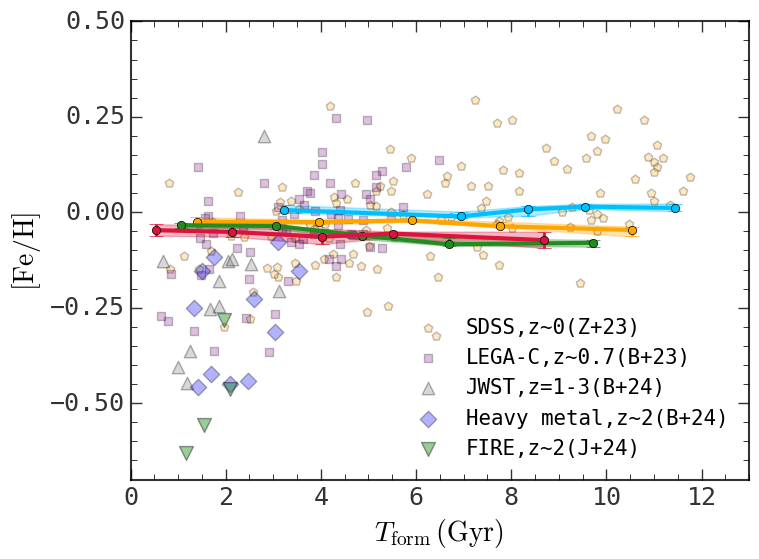}
    \caption{(Relative) Element abundances as a function of $T_{\rm form}$, binned by $\sigma_*$ (in km/s). Data points follow the same legend as Figure \ref{Fig8}. Our results in [Fe/H]-$T_{\rm form}$ plane are consistent with previous studies at $z \sim 0$ and $z \sim 0.7$ \citep[e.g.,][]{2023ApJ...948..132Z,2023ApJ...948..140B}. At higher redshifts \citep{2024arXiv240702556B,2024ApJ...966..234B,2024arXiv240603549J}, a decline in [Fe/H] at lower $T_{\rm form}$ suggests shifts in chemical evolution processes across cosmic time.}
    \label{Fig11}    
\end{figure*}

To further investigate whether and how $T_{\rm form}$ play a secondary role in determining %the relationship between $\sigma_*$ and 
element abundances, we conduct an analysis based on $\sigma_*$ binning-scheme. As shown in Figure~\ref{Fig11} and ~\ref{Fig8}, [Na/Fe], [Mg/Fe], [C/Fe], and [N/Fe] exhibit clear variations in abundance across different $\sigma_*$ bins, with [Na/Fe] demonstrating the most stable relationship with $T_{\rm form}$, largely unaffected by formation time. In contrast, [Mg/Fe], [C/Fe], and [N/Fe] show a mild anti-correlation with $T_{\rm form}$, especially at lower $\sigma_*$ values, where this trend is more pronounced.

In comparison, [Ca/Fe] and [Fe/H] display minimal variation in abundance with respect to both $\sigma_*$ and $T_{\rm form}$, with nearly identical abundance levels across all bins. This consistency for Ca and Fe across $\sigma_*$ values suggests that their enrichment processes are less sensitive to changes in gravitational potential or formation time, distinguishing them from the trends observed in other elements.

%\citep[e.g.,][]{2014ApJ...780...33C, 2023ApJ...948..132Z}, Massive, $z \sim 0$ \citep{2022ApJ...932..103G}, LEGA-C, $z \sim 0.7$ \citep{2023ApJ...948..140B}, JWST, $z \sim 1-3$ \citep{2024arXiv240702556B}, Heavy Metal, $z \sim 2$ \citep{2024ApJ...966..234B}, and FIRE, $z=2$ \citep{2024arXiv240603549J}.}

Figure~\ref{Fig10}, Figure~\ref{Fig11}, and the third column of Figure~\ref{Fig7} and~\ref{Fig8} also include the data points from various surveys and simulations among various redshift ranges for comparison \citep{2014ApJ...780...33C, 2022ApJ...932..103G, 2023ApJ...948..140B, 2023ApJ...948..132Z, 2024ApJ...966..234B, 2024arXiv240702556B, 2024arXiv240603549J}. Comparing these datasets with our results from Figures~\ref{Fig7} and \ref{Fig10}, we find that our data matches well with both $z \sim 0$ and $z \sim 0.7$ data for most elements, thereby confirming the robustness of our analysis. Furthermore, the plots with $T_{\rm form}$ on the x-axis suggest that data points from different low-redshift surveys may correspond to different $\sigma_*$ bins. This indicates that the high dispersion observed in these surveys likely arises from entirely distinct galaxy populations.

Additionally, we include data from \cite{Thomas2010,2012JohanssonThomas+,2014WortheyTang+,2021ParikhThomas+} for comparison, as these studies provide key insights into the mass-metallicity relation and element abundance trends in early-type galaxies. Though the data from \cite{Thomas2010} is for [$\alpha$/Fe], it still offers valuable information for comparison with [Mg/Fe], since Mg is one of the most representative elements among the $\alpha$-elements.  \cite{2012JohanssonThomas+,2021ParikhThomas+} used SSP model fitting, with \cite{2021ParikhThomas+} extending the analysis to full-spectrum fitting, yield similar values for [Na/Fe], [Mg/Fe], [N/Fe], and [Ca/Fe], but slightly higher values for [C/Fe] compared to our data. Focusing on absorption-line analysis, \cite{Thomas2010} and \cite{2014WortheyTang+} observed nuanced differences among elements, producing similar values for [Na/Fe] and [Mg/Fe], but slightly lower values for [C/Fe] and [N/Fe]. Regarding [Ca/Fe],  \cite{2014WortheyTang+} found an unexpectedly downward trend, despite abundance levels similar to our results and other literature. This decreasing trend could be caused by the model artefact. Overall, our data exhibit generally similar trends and
abundance levels compared to these studies, with a higher degree of approximation relative to
those using SSP model fitting \citep[e.g.][]{2012JohanssonThomas+,2021ParikhThomas+,2023ApJ...948..140B,2023ApJ...948..132Z}.

However, for higher redshifts, specifically in the JWST and Heavy Metal datasets, we observe some notable differences. The [Mg/Fe] at $z \sim 1-3$ is significantly higher than in our local sample, indicating enhanced magnesium production at earlier cosmic times. Conversely, [Fe/H] values at higher redshifts are lower compared to our results, suggesting delayed iron enrichment in these earlier epochs. %These discrepancies between high-redshift and local measurements might underscore the evolving nature of element abundances with cosmic time and/or the effect of minor mergers in the evolution of quenched galaxies. 
These discrepancies between high-redshift and local measurements might underscore the evolving nature of element abundances with cosmic time, reflecting processes such as enhanced magnesium production at earlier epochs or delayed iron enrichment. Alternatively, the observed high [Mg/Fe] at high redshifts could be influenced by modeling effects. For example, differences in spectral fitting techniques or assumptions in chemical evolution models may lead to systematic overestimations of [Mg/Fe], as noted in previous studies \citep[e.g.,][]{2005ArnoneRyan+,2019BlancatoNess+}.
%suggest that the enrichment histories of Mg and Fe are subject to different time scales and processes. 
Further investigation into these trends is required to better understand the relationship between element abundances and redshift.

\section{Discussion} \label{sec:4}
\subsection{$\sigma_*$ plays a primary role}

\begin{figure*}[htbp]
    \centering
    \includegraphics[width=1\textwidth]{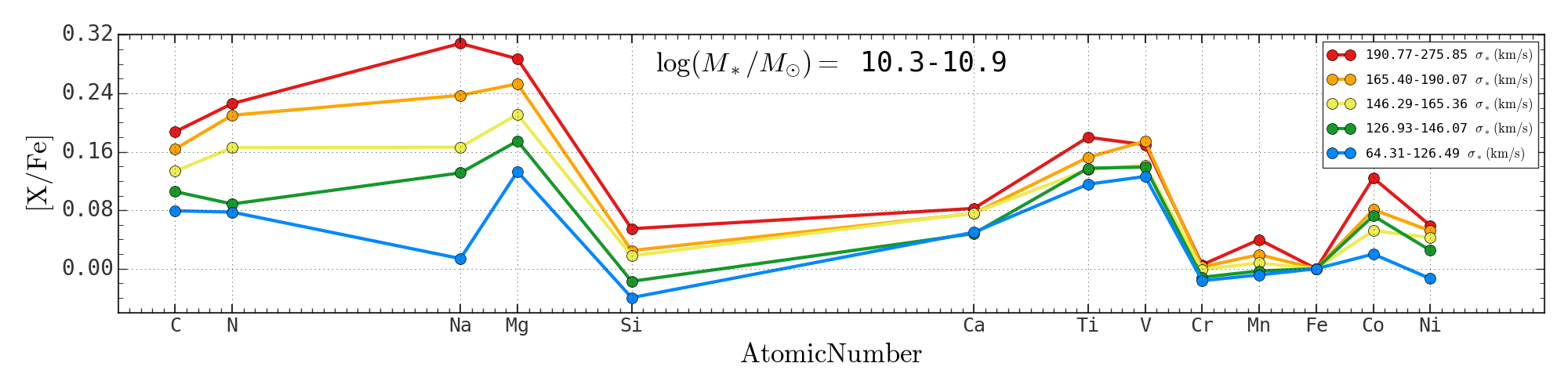}
    \includegraphics[width=1\textwidth]{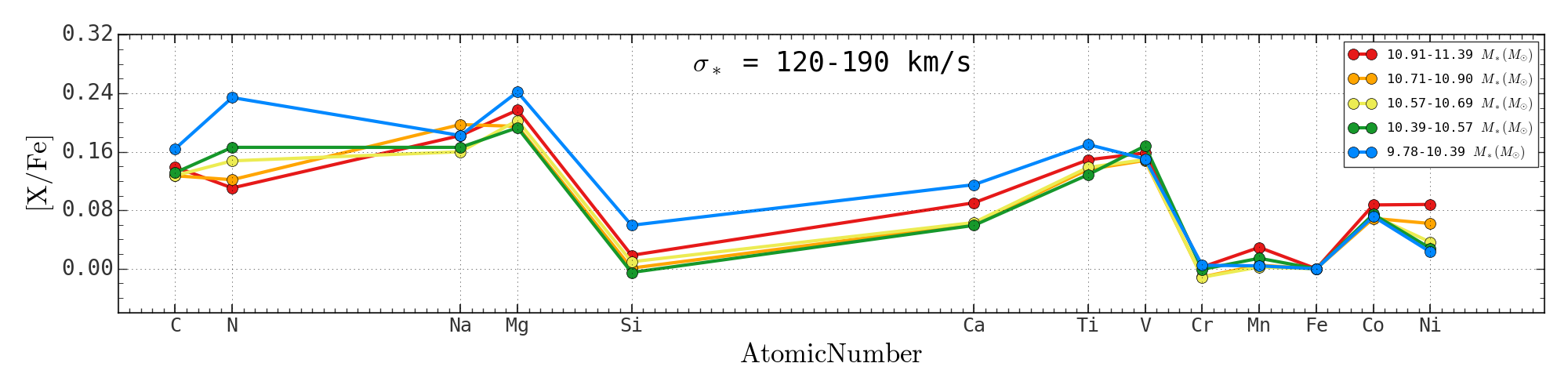}
    \includegraphics[width=1\textwidth]{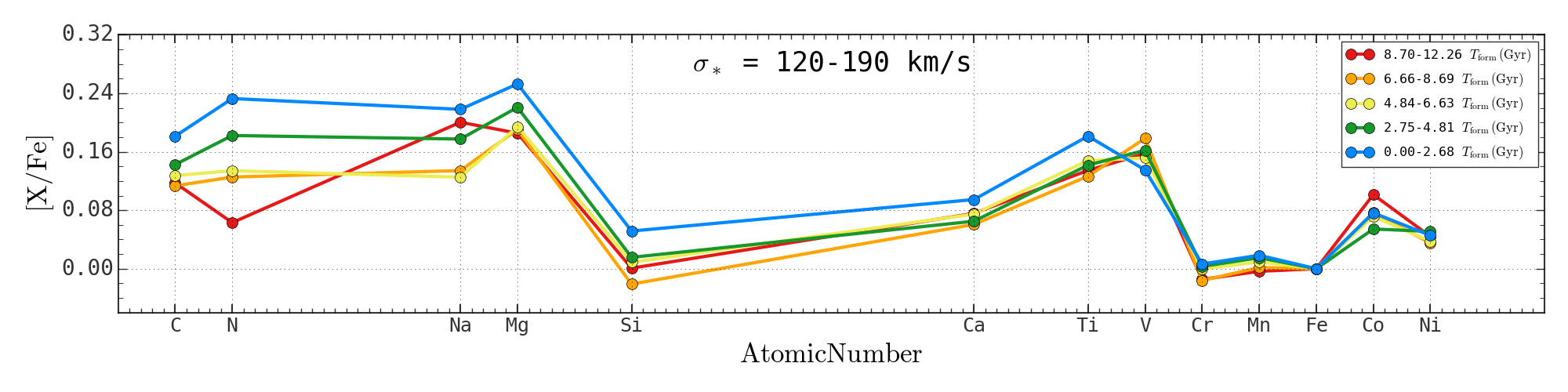}
    \caption{Relative element abundances [X/Fe] binned by velocity dispersion ($\sigma_*$, top panel), stellar mass ($M_*$, middle panel), and formation time ($T_{\rm form}$, bottom panel), respectively. The top panel shows 397 galaxies in $\log(M_*/M_{\odot})=$ 10.3-10.9, binned by $\sigma_*$. The median panel corresponds to 390 galaxies with $\sigma_*=$ 120-190 km/s, binned by $M_*$. The bottom panel contains 390 galaxies with $\sigma_*=$ 120-190 km/s, binned by $T_{\rm form}$, highlighting a slight effect on the overall chemical profile.}
    \label{Fig12}
\end{figure*}

The key findings from the analysis of (relative) element abundances for 13 elements across different binning methods are visually summarized in Figure~\ref{Fig12}. In the top and second panels, the galaxies are binned by $\sigma_*$ and $M_*$. Despite these similar mass and velocity dispersion ranges, the distributions exhibit significant differences. The top panel shows a clear trend where higher $\sigma_*$ values correlate with higher relative abundances of elements, such as [C/Fe], [N/Fe], [Na/Fe], and [Mg/Fe]. This indicates that velocity dispersion has a strong influence on chemical enrichment, particularly for elements tied to stellar processes that operate on different timescales. In contrast, as shown in the middle panel of Figure~\ref{Fig12}, there is no clear correlation between stellar mass and abundance patterns. This further supports the conclusion that velocity dispersion plays a key role in determining central abundance patterns.

In the third panel, galaxies are binned by $T_{\rm form}$, where lower $T_{\rm form}$ values (blue line) correspond to higher relative element abundances for elements like [C/Fe] and [N/Fe], which are produced over longer timescales. This suggests that $T_{\rm form}$ slightly affects the enrichment of elements, especially those with longer stellar evolution processes.
 
Our results highlight several key insights into the chemical enrichment processes within galaxies, with a particular focus on the relative element abundances of [Na/Fe], [Mg/Fe], [C/Fe], and [N/Fe], and their relationship with $\sigma_*$. %In this section, we discuss the implications of our findings and compare them to previous studies on galaxy metallicity and formation history. 
Specifically, the abundances of [Na/Fe], [Mg/Fe], [C/Fe], and [N/Fe] remain remarkably stable across different $T_{\rm form}$ bins when galaxies are binned by $\sigma_*$. This stability suggests that, for galaxies with the same $\sigma_*$, the element abundances do not vary significantly with their stellar age. Na and Mg are typically produced in relatively short-term core-collapse supernovae of massive stars, while C and N have more complex origins, involving both massive stars and dying low-mass stars (such as AGB stars). Thus, the production of these elements may be related to the burstiness of star formation during the early universe for local quenched galaxies. Although different $T_{\rm form}$ indicates relatively long-term evolution, it plays a secondary role compared to $\sigma_*$ for galaxies quenched at early times. Therefore, this finding suggests an underlying mechanism that may govern the star formation timescales of galaxies with similar central velocity dispersion.

%\citet{2018MNRAS.479.1807D} proposed aperture-matched subsampling, wherein galaxies are selected with physical sizes matched to the SDSS fiber aperture in units of effective radius ($R_\mathrm{e}$). Their work demonstrated that gas-phase metallicity correlates more tightly with the gravitational potential $(M_*/R_{\rm{e}})$ compared to stellar mass ($M_*$) or surface mass density $(M_*/R^2_{\rm{e}})$, suggesting that potential may be a significant factor in regulating metallicity. While this method reduces aperture bias in traditional single-fiber data, recent advancements in integral field spectroscopy, such as MaNGA \citep[Mapping Nearby Galaxies at Apache Point Observatory;][]{2015ApJ...798....7B} survey, provide a more direct solution. For instance, \citet{2024ApJ...971L..14M} u

%Several studies have highlighted the critical role of central stellar velocity dispersion ($\sigma_*$) in determining the metallicity and star formation histories of galaxies. For example, \citet{2014ApJ...780...33C} and \citet{2019ApJ...874...66G} proposed that $\sigma_*$ may be a more reliable indicator of a galaxy's chemical properties than stellar mass. Our findings further support this view by showing that $\sigma_*$ correlates more strongly with element abundances, such as [Fe/H] and [Mg/Fe], than either $M_*$ or $M_*/R_\mathrm{e}$, remaining a consistent indicator across different stellar ages. This suggests that $\sigma_*$ plays a central role in regulating the chemical evolution of galaxies.

Several studies have highlighted the critical role of gravitational potential (traced by $M_*/R_{\rm e}$) in determining the metallicity for both star-forming galaxies and quenched galaxies \citep[e.g.,][]{2018MNRAS.479.1807D, Vaughan2022, 2024ApJ...971L..14M}. 
%For example, \citet{2014ApJ...780...33C} and \citet{2019ApJ...874...66G} proposed that $\sigma_*$ may be a more reliable indicator of a galaxy's chemical properties than stellar mass. 
However, our findings support that, for quenched galaxies, $\sigma_*$ correlates more strongly with element abundances, such as [Fe/H] and [Mg/Fe], than either $M_*$ or $M_*/R_\mathrm{e}$, remaining a consistent indicator across different stellar ages. In consistent with this, \citet{2024MNRAS.534...30B} demonstrated that in passive galaxies, stellar metallicity predominantly correlates with $\sigma_*$,  suggesting that $\sigma_*$ plays a central role in regulating the chemical evolution of galaxies.

%Using MaNGA data, \citet{2024MNRAS.534...30B} demonstrated that in passive galaxies, stellar metallicity predominantly correlates with $\sigma_*$. Their analysis suggests that the $\sigma_*$ZR (velocity dispersion-metallicity relation) is not merely a reflection of dynamical mass or gravitational potential but is closely tied to black hole mass. They argued that this relationship arises from black hole feedback, wherein the black hole quenches star formation by heating the surrounding halo gas through either preventative or delayed feedback. This process inhibits gas accretion, leading to a state of "starvation" that halts star formation and influences the chemical properties of the galaxy, which is confirmed with \citet{2022ZhangWang+}, which also states the similar point that massive galaxies deplete their halo gas reservoirs through star formation, impacting long-term evolution.

\subsection{Possible explanations of the primary role of $\sigma_*$}
\subsubsection{Related to gravitational potential and dark matter halo}

The central velocity dispersion of galaxies is found to be more strongly connected to the color and clustering of quenched galaxies \citep{2012WakevanDokkum+, Wake2012b} than to stellar mass. 
Interestingly, \citet{2022ZhangWang+} found a similar thing that central velocity dispersion of galaxies is a better indicator of host dark matter halo mass than stellar mass at the high-mass end, using the weak lensing method. This motivates us to connect $\sigma_*$ to the assembly of dark matter halo mass in our interpretation of the results. 

The central velocity dispersion of galaxies has been found to have a stronger correlation with the color and clustering of quenched galaxies \citep{2012WakevanDokkum+, Wake2012b} than with stellar mass. Interestingly, \citet{2022ZhangWang+} also found that central velocity dispersion is a better indicator of host dark matter halo mass than stellar mass at the high-mass end, using the weak lensing method. This motivates us to relate $\sigma_*$ to the assembly history of dark matter halo mass in our interpretation of the results.

Specifically, \citet{2024MNRAS.532.3808M} proposed that the build-up of $\sigma_*$ and its strong correlation 
with element abundances is the two-phase scenario of galaxy formation. 
This scenario is based on the fact that dark matter halos in the $\Lambda$CDM cosmology show a universal assembly pattern 
consisting of a phase of fast accretion at high $z$, where the gravitational 
potential rapidly deepens, followed by a phase of slow accretion at low $z$, 
where the potential well stabilizes, and the accreted matter accumulates 
in the outskirts. During the fast phase, the large gas fraction due to 
effective cooling, combined with the frequent perturbation of potential 
by accretion, leads to the fragmentation of self-gravitating gas cloud by 
Jeans instability into dense, ballistically moving sub-clouds before the gas 
can settle into a dynamically cool disk. Star clusters formed in these sub-clouds inherit their dynamic hotness, with tidal heating disrupting them later, forming a stellar bulge dominated by random motion. A thin disk appears later only if the halo transitions into the slow phase and if feedback has not yet depleted the gas reservoir. Such a scenario predicts an in-situ origin of the inner stellar content of a quenched galaxy and motivates us to restrict the analysis within $0.5 R_{\rm e}$.

The link between fast halo assembly and dynamic hotness provides clues into the 
fundamental role of $\sigma_*$ in determining element abundance, as 
found in this work. Firstly, halos with higher mass not only have deeper inner 
potentials, which more effectively trap the enriched outflow produced 
by star formation, but also have earlier assembly times, leading to a shorter 
dynamical timescale that promotes bursty star formation. 
Secondly, at a given virial mass, halos with earlier assembly times have 
larger concentration and thus deeper inner potential, which again aids 
in trapping the enriched outflow and promotes bursty star formation. 
Both factors lead to a positive correlation between inner potential 
as reflected by $\sigma_*$, and star-formation burstiness, as indicated by 
$\alpha$-enhancement, regardless of whether $M_*$ is fixed or not, thus 
explaining the predictive capability of $\sigma_*$ over $M_*$ \citep[also see][]{2024ChenMo+}.

\subsubsection{Related to black hole and feedback}

Another explanation for these trends is that central velocity dispersion of quenched galaxies is a key observable that correlates strongly with mass of central supermassive black hole, making it a reliable tracer of black hole activity \citep[e.g.,][]{2013ARA&A..51..511K,2016ApJ...818...47S,2022MNRAS.512.1052P,2024ParkBelli+}. $\sigma_*$ also encodes the integrated history of feedback associated with black hole accretion. In particular, AGN feedback, through either preventative or ejective modes, suppresses star formation and transforms galaxies from star-forming to quenched states \citep[e.g.,][]{2016MNRAS.462.2559B,2020MNRAS.491.2939O,2020MNRAS.499..230B,2024MNRAS.534...30B}. This AGN feedback prevents gas from cooling and fuels the quenching process, fundamentally alters both the star formation activity and kinematic structure of the galaxy.

In particular, \citet{2024MNRAS.534...30B} suggested that the velocity dispersion-metallicity relation is not merely a reflection of dynamical mass or gravitational potential but is closely tied to black hole mass. They argued that this relationship arises from black hole feedback, wherein the black hole quenches star formation by heating the surrounding halo gas through either preventative or delayed feedback. This process inhibits gas accretion, leading to a state of ``starvation'' that halts star formation and influences the chemical properties of the galaxy. 
 %, which is confirmed with \citet{2022ZhangWang+}, which also states the similar point that massive galaxies deplete their halo gas reservoirs through star formation, impacting long-term evolution.

%While our results provide strong evidence for a relationship between $\sigma_*$ and galaxy abundance patterns, further studies are necessary to confirm these trends, especially by focusing on the detailed star formation histories of galaxies with comparable $\sigma_*$. Additionally, investigating other environmental factors, such as local galaxy density and merger history, may reveal additional variables that influence these abundance patterns. Such factors are particularly important in environments with frequent interactions, which could have significant effects on both chemical enrichment and star formation. These studies are essential for clarifying the broader role of $\sigma_*$ in galaxy growth and chemical evolution.

In conclusion, our findings advance the understanding of how central velocity dispersion influences galactic chemical evolution and star formation history. Future research will need to disentangle the complex interactions between dark matter, baryonic processes, and galaxy evolution by incorporating more sophisticated models that account for dark matter halo properties, merger histories, and environmental conditions, thus offering a more complete picture of how these factors shape star formation and chemical enrichment.

\section{Summary} \label{sec:5}
In this work, we analyze high-resolution IFU data from the MaNGA survey, focusing on a sample of 1,185 quenched galaxies spanning redshifts from 0.012 to 0.15. Our primary objective is to investigate the relationships between various element abundances and key galactic parameters, with particular emphasis on central velocity dispersion, stellar mass, and mass-to-effective radius ratio. The summary of our results is listed below: 

\begin{enumerate}
    
    \item \textbf{Significance of $\sigma_*$:} Our analysis identifies $\sigma_*$ as the most effective parameter for correlating with relative element abundances, especially for [Na/Fe], [Mg/Fe], [C/Fe], and [N/Fe]. The stronger correlations suggest that galaxies with similar $\sigma_*$ likely share comparable formation histories and star formation burst patterns. Thus, $\sigma_*$ serves as a more consistent and predictive measure of element abundance patterns compared to other mass-related parameters, such as $M_*$ and $M_*/R_\mathrm{e}$, underscoring its central role in tracing galaxy evolution.
    
    \item \textbf{Element Abundances and $\sigma_*$ Stability:} When binned by $\sigma_*$, the abundances of [Na/Fe], [Mg/Fe], [C/Fe], and [N/Fe] display notable stability across different $T_{\rm form}$ bins, suggesting that age variations have limited impact on these abundances. This stability implies that $\sigma_*$ provides a clearer and more direct representation of a galaxy's chemical evolution, largely unaffected by external factors, such as merger events, and suggests that galaxies with similar $\sigma_*$ may have experienced comparable evolutionary processes.
    
    \item \textbf{Contribution of $T_{\rm form}$ to Element Abundances:} Although $\sigma_*$ proves to be a dominant factor, our analysis also reveals a meaningful contribution from $T_{\rm form}$ to element abundances. $T_{\rm form}$ impacts the long-term evolution of elements, particularly Fe and Ca, which appear more closely tied to steady, prolonged star formation rather than the short-lived, intense bursts linked to $\sigma_*$. This indicates that, while $\sigma_*$ reflects the rapid star formation episodes critical for elements like Na, Mg, C, and N, $T_{\rm form}$ plays a complementary role in the secular enrichment of elements associated with longer evolutionary timescales.

\end{enumerate}

Overall, our work offers a more nuanced understanding of the interplay between galactic parameters and element abundances. By emphasizing the importance of $\sigma_*$ in determining chemical properties and minimizing external contamination, we contribute to the broader understanding of galaxy formation history and chemical evolution. Future research should further explore these relationships, particularly the potential role of dark matter halos and star formation histories in shaping the observed metallicity trends.

%% IMPORTANT! The old "\acknowledgment" command has be depreciated. It was
%% not robust enough to handle our new dual anonymous review requirements and
%% thus been replaced with the acknowledgment environment. If you try to 
%% compile with \acknowledgment you will get an error print to the screen
%% and in the compiled pdf.
%% 
%% Also note that the akcnowlodgment environment does not support long amounts of text. If you have a lot of people and institutions to acknowledge, do not use this command. Instead, create a new 
\section{Acknowledgments}.
\begin{acknowledgments}

The authors thank the anonymous referee for their helpful comments that improved the quality of this paper. EW thanks support of the National Science Foundation of China (Nos. 12473008). HYW is supported by the National Natural Science Foundation of China (Nos. 12192224) and CAS Project for Young Scientists in Basic Research, Grant No. YSBR-062. The authors also thank Charlie Conroy for providing the {\tt alf} model and for his early support, as well as Aliza G. Beverage for sharing data from various sources and for insightful discussions on high-redshift literature. The authors gratefully acknowledge the support of Cyrus Chun Ying Tang Foundations. 
\end{acknowledgments}

\bibliography{sample631}{}
\bibliographystyle{aasjournal}
\newpage
\appendix
\section{The relations between galaxy properties and the other six elements in hexbin plots}
\begin{figure}[h!]
    \centering
    \includegraphics[width=0.75\textwidth]{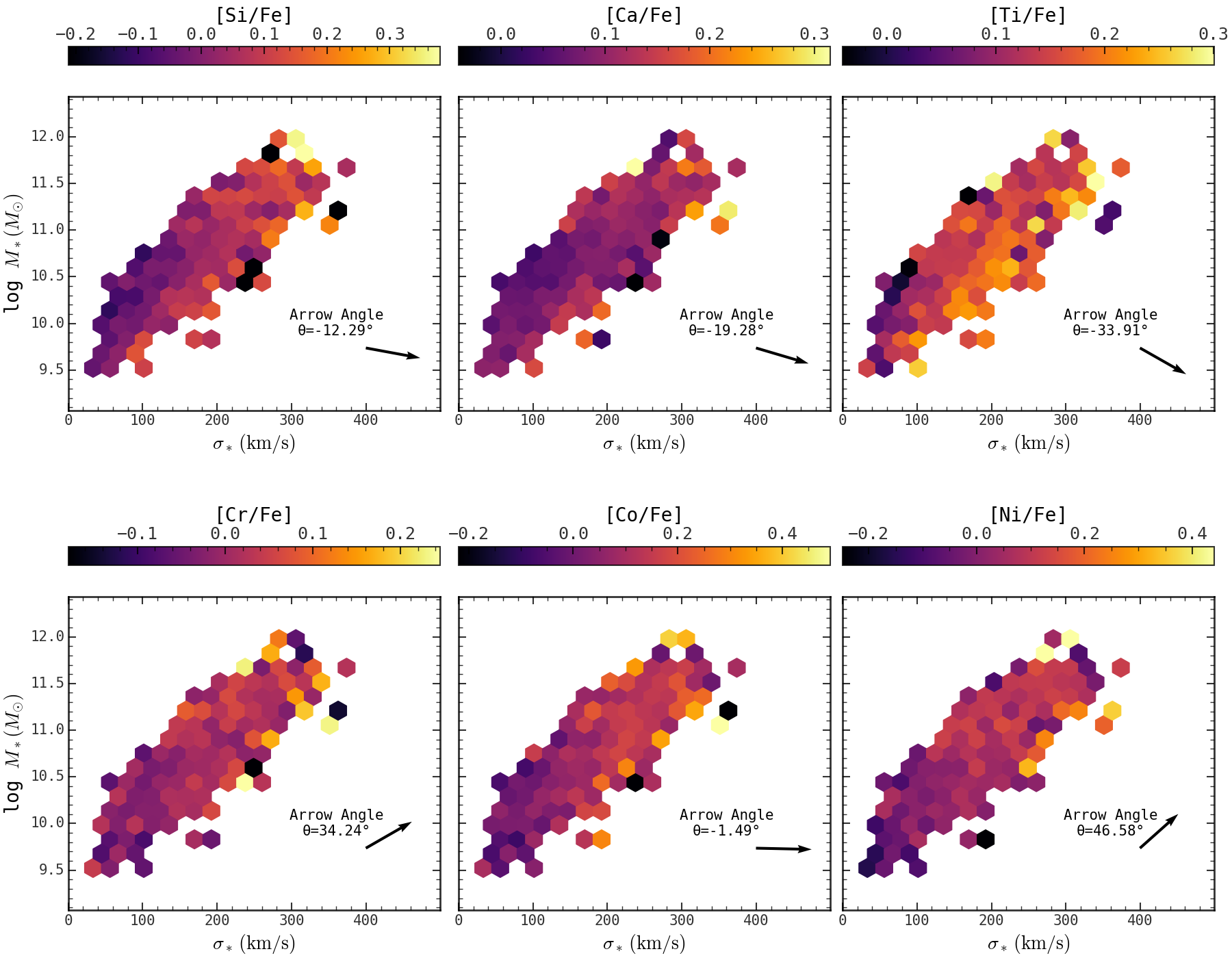}
    \caption{Relative element abundances [Si/Fe], [Ca/Fe], [Ti/Fe], [Cr/Fe], [Co/Fe], and [Ni/Fe] as functions of \(\sigma_*\) and \(M_*\). The color gradients, indicated by arrows, represent the directions of increasing abundances.}
    \label{Fig13}
\end{figure}

\begin{figure}
    \centering
    \includegraphics[width=0.75\textwidth]{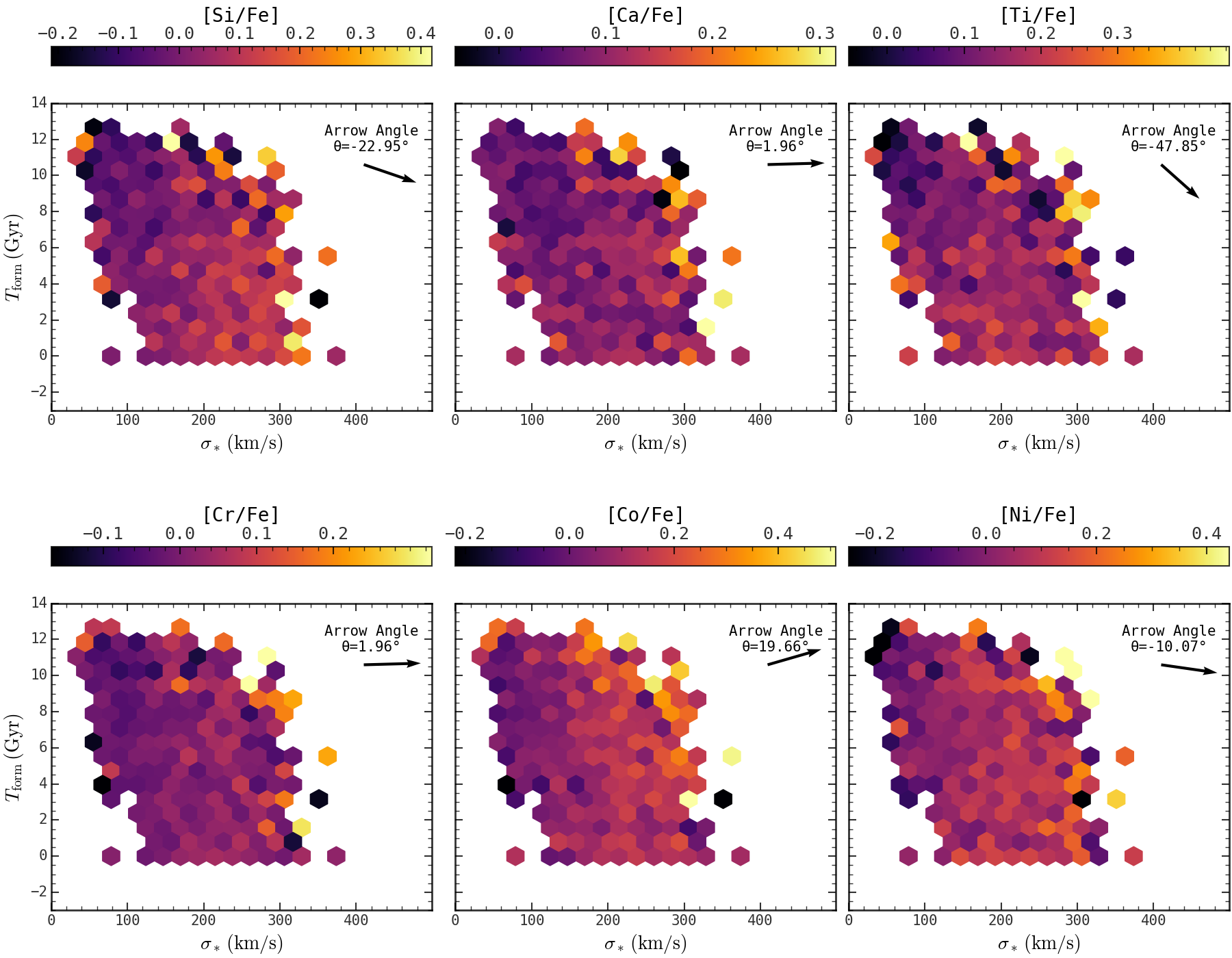}
    \caption{Similar to Figures~\ref{Fig13} but for relative element abundances as functions of \(\sigma_*\) and \(T_{\text{form}}\).}
    \label{Fig14}
\end{figure}

\begin{figure}
    \centering
    \includegraphics[width=0.75\textwidth]{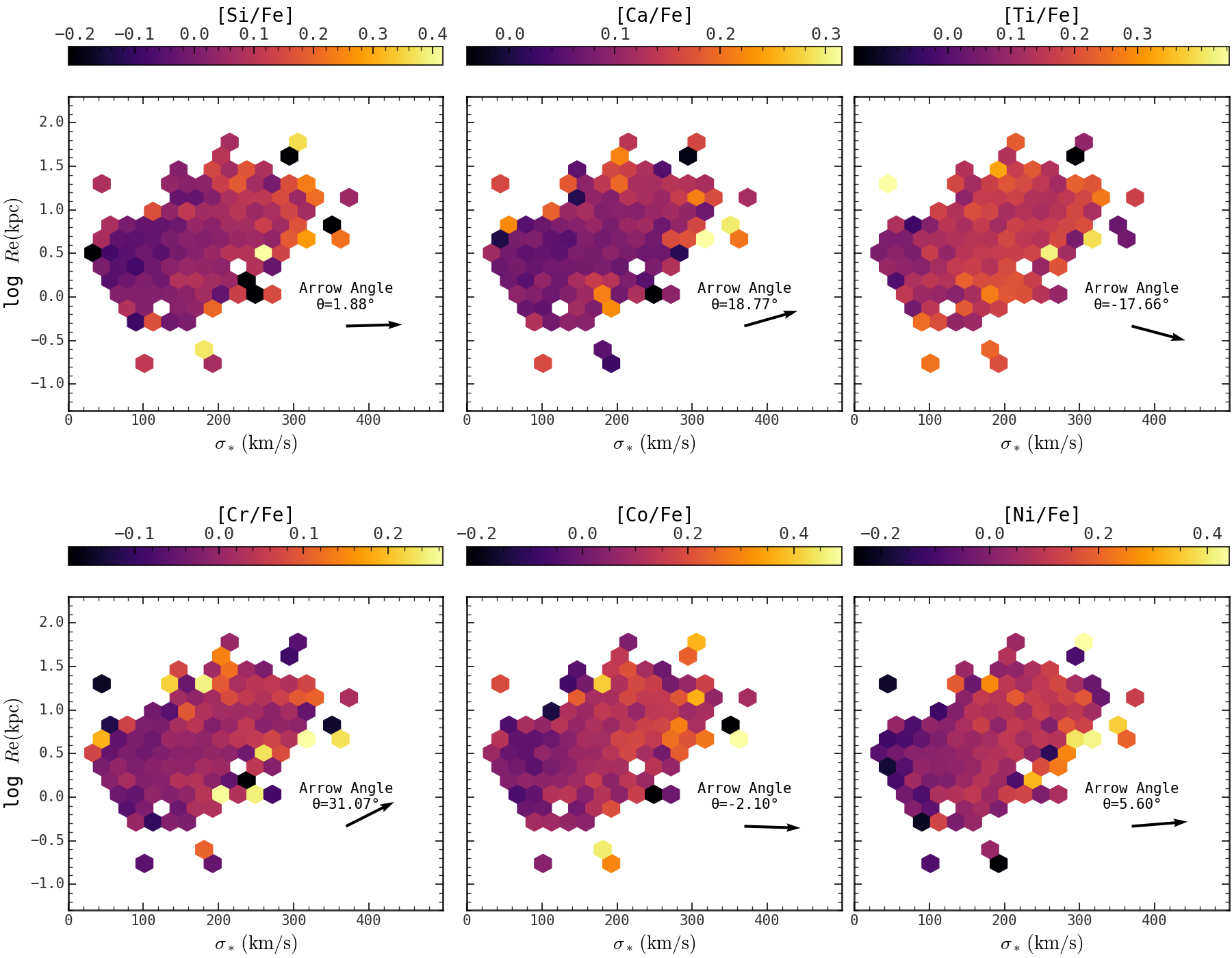}
    \caption{Similar to Figures~\ref{Fig13} but for relative element abundances as functions of \(\sigma_*\) and \(R_\mathrm{e}\).}
    \label{Fig15}
\end{figure}

Figures~\ref{Fig13}, \ref{Fig14}, and \ref{Fig15} illustrate additional relative element abundances from the \texttt{alf} fitting for [Si/Fe], [Ca/Fe], [Ti/Fe], [Cr/Fe], [Co/Fe], and [Ni/Fe]. Similar to Figures~\ref{Fig5} and \ref{Fig6}, these figures use $\sigma_*$ as x-axis, with Figure~\ref{Fig13} displaying $M_*$ as y-axis, Figure~\ref{Fig14} showing $T_{\rm form}$, and Figure~\ref{Fig15} representing $R_e$, respectively. The color gradients, indicated by arrows, mark the directions of increasing element abundance or age. These six elements notably show subtle color gradient distinctions, particularly for [Ca/Fe] and [Cr/Fe].

All three figures demonstrate a strong correlation between these six elements and $\sigma_*$ due to the arrow angles. Specifically, [Ti/Fe], [Cr/Fe], and [Ni/Fe] show significant correlation with $M_*$ (angles all exceed 30°). Notably, [Ni/Fe] exhibits the strongest relationship with $M_*$, as its relative abundance correlates more positively with $M_*$ than with $\sigma_*$, resulting in an angle of over 45°. [Ti/Fe] positively correlates with $M_*$, while [Cr/Fe] displays a notable inverse relationship.

When considering $T_{\rm form}$ (Figures~\ref{Fig14}) and $R_e$ (Figures~\ref{Fig15}) as y-axis, only [Ti/Fe] shows a strong inverse correlation with $T_{\rm form}$, and [Cr/Fe] displays a positive correlation with $R_e$. The inverse correlation between [Ti/Fe] and $T_{\rm form}$ is even stronger than its correlation with $\sigma_*$, resulting in an angle of approximately -47.85°. However, given the subtle color distinctions in these elements, especially for [Ca/Fe] and [Cr/Fe], the accuracy of our gradient angle measurements warrants further verification.

\newpage
\section{The relations between galaxy properties and the other three elements in median plots}
\begin{figure}
    \centering
    \includegraphics[width=0.8\textwidth]{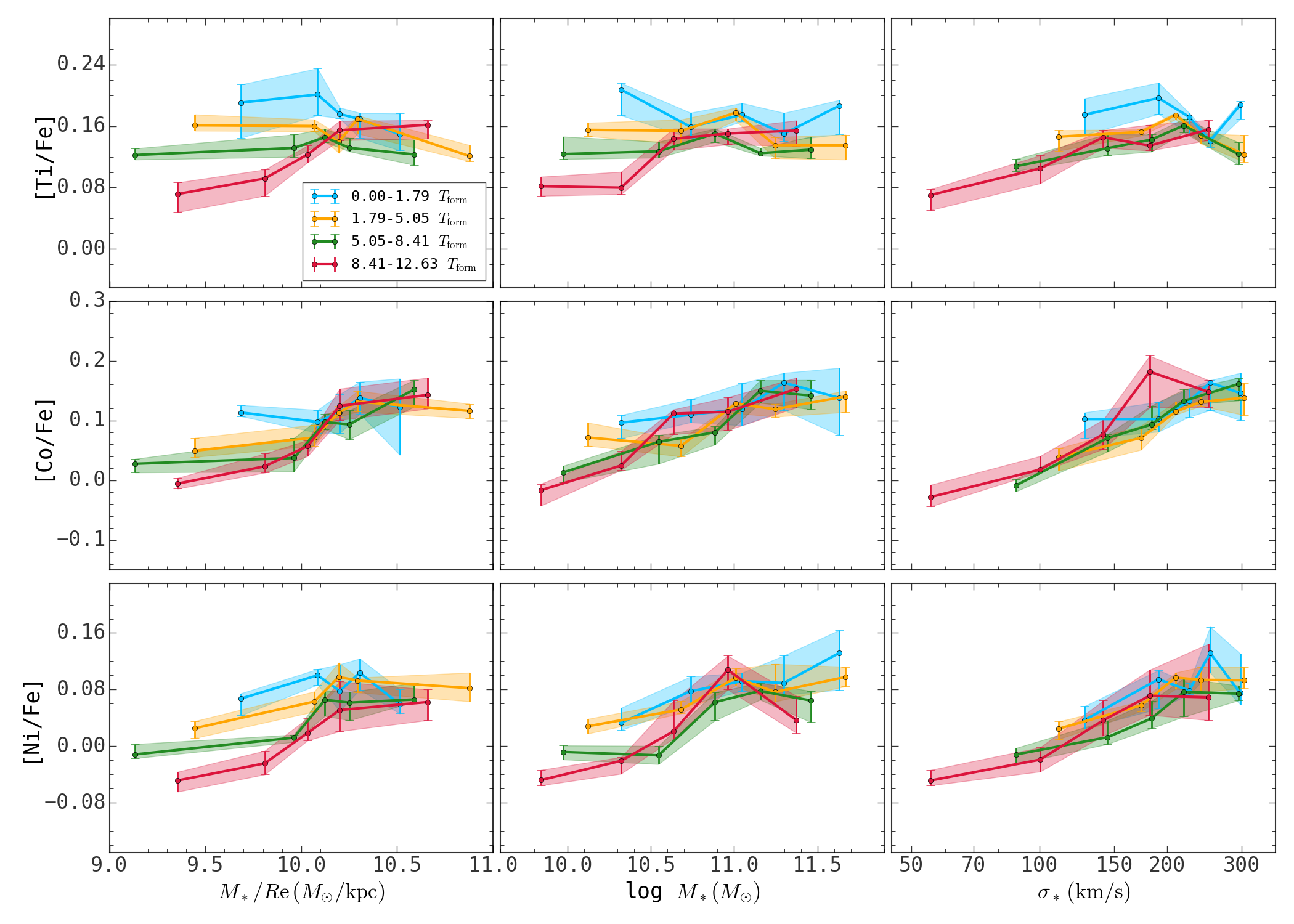}
    \caption{Similar to Figures~\ref{Fig7} but for relative element abundances [Ti/Fe], [Co/Fe], and [Ni/Fe].}
    \label{Fig16}
\end{figure}

\begin{figure}
    \centering
    \includegraphics[width=0.8\textwidth]{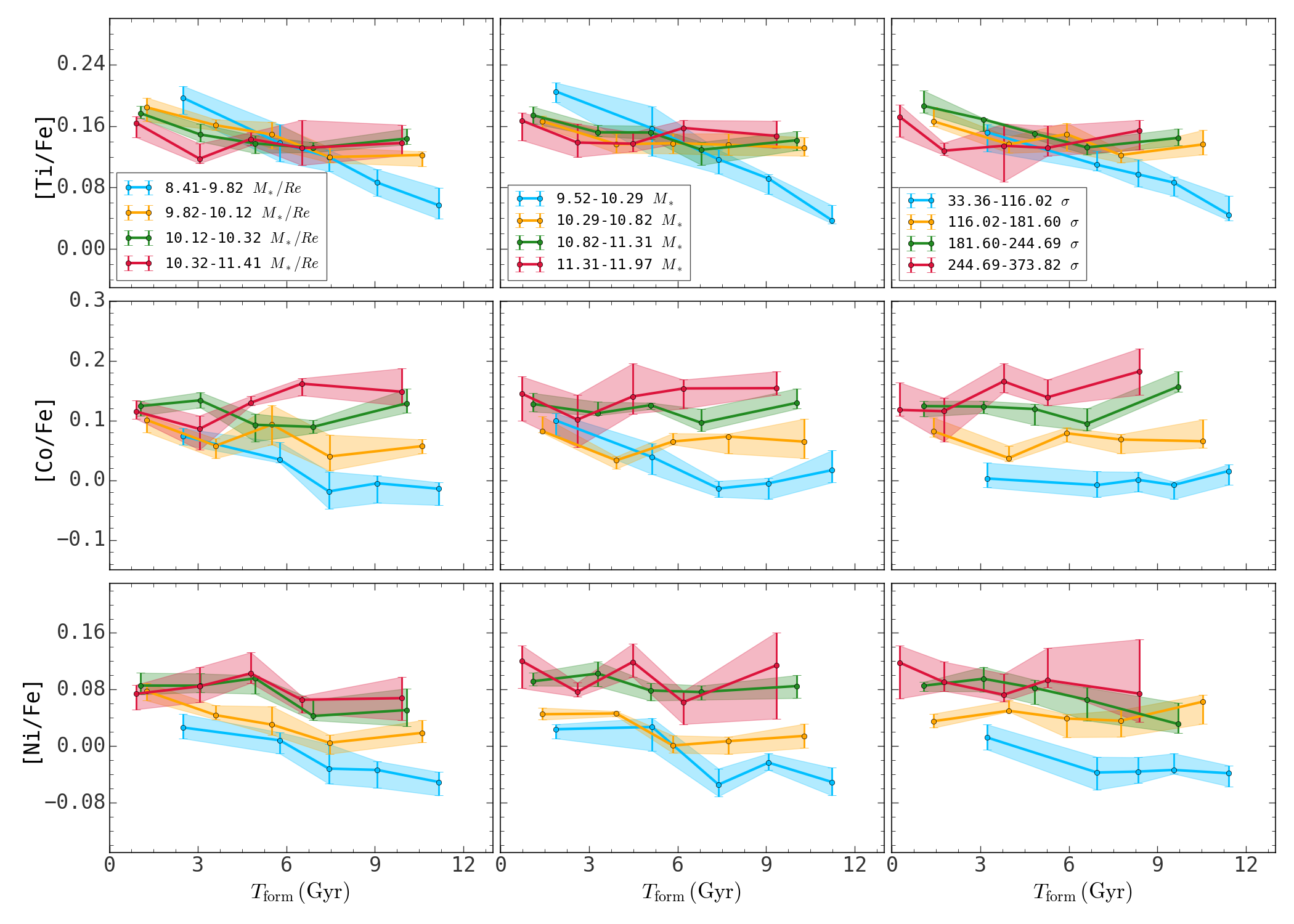}
    \caption{Similar to Figures~\ref{Fig8} but for relative element abundances [Ti/Fe], [Co/Fe], and [Ni/Fe].}
    \label{Fig17}
\end{figure}

Figures~\ref{Fig16} and \ref{Fig17} present median line plots for [Ti/Fe], [Co/Fe], and [Ni/Fe], similar to Figures~\ref{Fig7} and \ref{Fig8}. In Figure~\ref{Fig16}, relative element abundances are shown as a function of $T_{\rm form}$, with the data binned by $M_*/R_e$ (left panels), $M_*$ (middle panels), and $\sigma_*$ (right panels). Figure~\ref{Fig17} displays these three relative element abundances as functions of $M_*/R_e$ (left panels, in $M_\odot$/kpc), $M_*$ (middle panels, in $M_\odot$), and $\sigma_*$ (right panels, in km/s), with the data binned by $T_{\rm form}$.

In Figure~\ref{Fig16}, the trend is similar to that in Figure~\ref{Fig7}: the $\sigma_*$ panels exhibit a slightly higher positive correlation, with multiple $T_{\rm form}$ bins showing strong consistency within the $\sigma_*$ column and aligning along an approximate sequence. This indicates that $\sigma_*$ has a stable influence on the relative abundances of [Ti/Fe], [Co/Fe], and [Ni/Fe] across different formation times.

In Figure~\ref{Fig17}, in addition to [Co/Fe] and [Ni/Fe], which exhibit similar behaviors to [Mg/Fe] as seen in previous figures, [Ti/Fe] shows nearly identical trends across the three panels. High mass or high $\sigma_*$ bins display consistent distributions, while lower mass or $\sigma_*$ bins reveal a pronounced negative correlation with $T_{\rm form}$. Combined with the high correlation observed between [Ti/Fe] and $M_*$ in Figures~\ref{Fig13}, \ref{Fig14}, and \ref{Fig15}, this suggests a unique origin for Ti elements, potentially indicating that Ti has distinct formation pathways compared to other elements.

%% This command is needed to show the entire author+affiliation list when
%% the collaboration and author truncation commands are used.  It has to
%% go at the end of the manuscript.
%\allauthors

%% Include this line if you are using the \added, \replaced, \deleted
%% commands to see a summary list of all changes at the end of the article.
%\listofchanges

\end{document}